\documentclass[preprints,article,accept,pdftex,moreauthors]{mdpi}

\usepackage{soul} \usepackage{xcolor}

\newcommand{\p}{\partial}
\newcommand{\rd}{\mathrm{d}}
\newcommand{\mG}{\mathcal{G}}

\newcommand{\mH}{\mathcal{H}}
\newcommand{\rA}{\mathrm{A}}
\newcommand{\rH}{\mathrm{H}}
\newcommand{\rT}{\mathrm{T}}
\newcommand{\rC}{\mathrm{C}}
\newcommand{\rtot}{\mathrm{tot}}
\newcommand{\deltaD}{\delta_{\mathrm{D}}}

\newcommand{\half}{\tfrac{1}{2}}
\newcommand{\bg}{\mathbf{g}}
\newcommand{\bu}{\mathbf{u}}
\newcommand{\bD}{\mathbf{\Delta}}
\newcommand{\eps}{\epsilon}
\newcommand{\cst}{\mathrm{cst}}
\newcommand{\kJ}{k_{\mathrm{J}}}
\newcommand{\kp}{k_{\mathrm{p}}}
\newcommand{\tr}{\tilde{r}}
\newcommand{\Ekin}{E_{\mathrm{kin}}}
\newcommand{\Egr}{E_{\mathrm{gr}}}
\newcommand{\br}{\mathbf{r}}
\newcommand{\btr}{\tilde{\mathbf{r}}}
\newcommand{\bp}{\mathbf{p}}
\newcommand{\bm}{\mathbf{m}}
\newcommand{\Max}{\mathrm{Max}}
\newcommand{\Tdyn}{T_{\mathrm{dyn}}}

\firstpage{1} 
\pubvolume{1}
\issuenum{1}
\articlenumber{0}
\pubyear{2025}
\copyrightyear{2025}

\Title{Thermodynamic blocking in self-gravitating systems}

\Author{Barnab\'{a}s Deme$^{1,2}$, Jean-Baptiste Fouvry$^{3}$}

\address{$^{1}$ \quad Baja Astronomical Observatory of SZTE University, Szegedi \'{u}t, Kt. 766, Hungary, H-6500 \\
$^{2}$ \quad ELTE E\"{o}tv\"{o}s Lor\'{a}nd University, Institute of Physics and
Astronomy, Department of Astronomy, P\'{a}zm\'{a}ny P\'{e}ter stny. 1/A,
H-1117, Budapest, Hungary\\
$^{3}$ \quad Institut d’Astrophysique de Paris, UMR 7095, 98 bis Boulevard Arago, F-75014 Paris, France}

\corres{deme.barnabas@ttk.elte.hu; fouvry@iap.fr}

\abstract{Building upon a thermodynamic formalism,
we show that self-gravitating systems in hydrostatic equilibrium
with a uniform density
are maximal entropy states when submitted to perturbations which are slow on dynamical timescale.
We coin this phenomenon "thermodynamic blocking",
given its similarity with the more general "kinetic blocking".
This result underlines the importance of the thermodynamic formalism
which proves useful when kinetic equations break down. }

\keyword{Kinetic theory; Self-gravitating systems; Relaxation; Entropy}

\begin{document}

\section{Introduction}

The statistical physics of self-gravitating systems is crucial to understand globular clusters, galaxies, or dark matter halos~\cite{binney2008,hamilton2024}. Therein, the number of particles, $N$, is a key parameter~\cite{campa2014}. Indeed, if ${ N \!\to\! + \infty }$, one is in the collisionless limit, i.e.\@, particles interact only with the background mean potential they create. Following a short and violent evolution~\cite{lyndenbell1967}, such systems are shown to settle down to stationary states which may be far from thermodynamic equilibrium.

If, on the other hand, ${ N \!<\! + \infty }$, then the system slowly evolves through stationary states to its thermodynamic equilibrium.\footnote{Three-dimensional self-gravitating systems with finite mass do not even have, stricto sensu, such a thermodynamic equilibrium. Equilibria are possible only if the system is enclosed in a finite box~\cite{chavanis2002,padmanabhan1989}.}
Such a relaxation is driven by Poisson fluctuations in the distribution function.
Assuming that the mean system is linearly stable and integrable,
this relaxation is described by the Balescu--Lenard equation~\cite{heyvaerts2010,Chavanis2012}.
This is the governing equation of relaxation in action space as being driven by resonant interactions between the particles.
Let us point out two remarks. First, in one-dimensional systems,
if the frequency profile is monotonic,
the ${ 1/N }$ Balescu--Lenard equation predicts a vanishing flux in action space.
This is called "kinetic blocking"~\cite[see, e.g.\@,][and references therein]{Eldridge+1963,Dawson+1964,Rouet+1991,Miller1996,Bouchet+2005,Rocha+2014,fouvry2023}.
Second, if the matter distribution is uniform (in one or higher dimensions),
then Poisson equation (Eq.~\ref{eq:poisson} below) predicts harmonic oscillations for the particles,
i.e. the orbital frequency is independent of the orbit.
Because it makes orbital resonances ubiquitous,
this latter property invalidates the Balescu--Lenard formalism.
As such, this kinetic equation cannot provide any information about the long-term evolution
of homogeneous self-gravitating systems.

In this work, we focus on this latter regime, namely homogeneous self-gravitating systems.
Building upon a thermodynamic formalism,
we show that, in these systems, matter flows (i.e.\ convective flows)
and heat flows (i.e.\ conductive flows) do not contribute to entropy growth.
Up to a particular approximation (see below),
we conclude that these flows are prohibited
in homogeneous self-gravitating systems.
This phenomenon is reminiscent of kinetic blocking.
On the one hand, it is more general because it works in higher dimensions as well.
On the other hand, it is less stringent than kinetic blocking, as it prohibits only macroscopic flows,
while kinetic blocking freezes microscopic flows (redistribution of actions) as well. Given these similarities, we coin this phenomenon "thermodynamic blocking".

The structure of the paper is as follows. In Section~\ref{sec:1d}, we focus on 1D self-gravitating systems.
In particular, we introduce our thermodynamic formalism (Section~\ref{sec:kinetic_formalism}),
detail how to perform thermodynamic variations (Section~\ref{sec:entropy_var}),
recover the expected thermodynamic equilibrium (Section~\ref{sec:Thermo_Eq}),
and finally uncover the phenomenon of thermodynamic blocking
in homogeneous systems (Section~\ref{sec:tb}).
In Section~\ref{sec:3d}, we detail how the same results
also hold in 3D self-gravitating systems. Finally, we briefly discuss our results in Section~\ref{sec:discussion}.

\section{1D self-gravitating systems}
\label{sec:1d}

One-dimensional self-gravitating systems are particularly useful and insightful models~\citep[see][for a throrough review]{Miller+2023}:
they may be used to model gravitational collapse during the formation
of cosmoslogical large-scale structures~\cite[see, e.g.\@,][]{Valageas2006,Schulz+2013};
to understand the intricacy of collisionless (violent) relaxation~\citep[see, e.g.\@,][]{Joyce+2011,teles2011,Colombi+2014}
and collisional (slow) relaxation~\citep[see, e.g.\@,][]{Joyce+2010,roule2022}
or even to describe the spontaneous thickening of galactic discs~\citep[see, e.g.\@,][]{Tremaine+2023}.
In practice, the modelling of 1D self-gravitating systems is made more tractable
owing to their reduced number of degrees of freedom.
As a result, here, we first describe thermodynamic blocking in the 1D case.
Most of the present results will also be valid in the 3D case (Section~\ref{sec:3d}).
We refer to~\cite{Padmanabhan1990,LyndenBell1999,Chavanis+2002,Katz2003}
for detailed reviews on some of the key aspects
of the thermodynamics of self-gravitating systems.

\subsection{From kinetics to thermodynamics}
\label{sec:kinetic_formalism}

Let us start with the formalism of kinetic theory in 1D~\citep[see][for some generic introduction]{binney2008}. We define ${ f(x,v,t) }$ as the distribution function in the phase space spun by the coordinate, $x$, and its canonically conjugate momentum, the velocity, ${ v \!=\! \rd x / \rd t }$.
We define $f$ such that its integral over the whole phase space is the total mass.

Assuming that the dynamics of the particles is driven by
the Hamiltonian 
\begin{equation}
\mH = \half v^{2} + \phi(x)
\end{equation}
with $\phi$ being the potential, then $f$ evolves according the collisionless Boltzmann/Vlasov equation~\citep{binney2008}
\begin{equation}
\frac{\p f}{\p t} + v \frac{\p f}{\p x} - \frac{\p \phi}{\p x} \frac{\p f}{\p v} = 0 ,
\label{eq:boltzmann}
\end{equation}
where we used Hamilton's canonical equations of motion.\footnote{A non-zero collision operator on the right-hand side of Eq.~\eqref{eq:boltzmann} would originate from pairwise interactions between particles.}

As usual, we now introduce the density and velocity moments with
\begin{subequations}
\begin{align}
\rho (x) {} & = \!\! \int \!\! \rd v \, f (x , v) ,
\label{eq:def_rho}
\\
\langle v^{n} \rangle (x) {} & = \frac{1}{\rho (x)} \!\! \int \!\! \rd v \, v^{n} \, f (x , v) .
\label{eq:def_vn}
\end{align}
\label{eq:def_moments}\end{subequations}
We can compute the velocity moments of Eq.~\eqref{eq:boltzmann}
to obtain the Jeans equations.
More precisely, combining the zeroth and first moments, one gets~\citep{binney2008}
\begin{equation}
\rho \big( \dot{\langle v \rangle} + \langle v \rangle \, \langle v \rangle' \big) = -\big[ \rho \, \big( \langle v^2 \rangle - \langle v \rangle^{2} \big) \big]' - \rho \, \phi' .
\label{eq:Jeans}
\end{equation}
In this exrepssion, and for the rest of this section,
we dropped the dependence with respect to $x$.
In Eq.~\eqref{eq:Jeans},
the dot (resp.\ the prime) denotes a partial derivative
with respect to $t$ (resp.\ $x$).
Of course, Eq.~\eqref{eq:Jeans} is not closed.
The dynamics of ${ \langle v \rangle }$
involves ${ \langle v^{2} \rangle }$, whose evolution equation itself involves
higher-order moments, and so on.

Progress can be made by placing ourselves in the time-stationary limit.
Let us therefore assume that there is neither time derivative (${ \dot{\langle v \rangle} \!=\! 0 }$)
nor bulk motion (${ \langle v \rangle \!=\! 0 }$).\footnote{For an isolated system,
one can always set ${ \langle v \rangle \!=\! 0 }$
by placing oneself in an appropriate inertial frame.}
Equation~\eqref{eq:Jeans} then reduces to the simple form
\begin{equation}
p' = \rho g ,
\label{eq:rho_T}
\end{equation}
where, once again, the dependence with respect to $x$ was dropped,
and we introduced the relevant quantities
\begin{subequations}
\begin{align}
g {} & = - \phi' && \text{(Gravitational acceleration)}
\label{eq:def_g}
\\
p {} & = \rho \, T && \text{(Pressure)}
\label{eq:def_p}
\\
T {} & = \langle v^{2} \rangle && \text{(Temperature)}
\label{eq:def_T}
\end{align}
\label{eq:many_def}\end{subequations}
Importantly, Eq.~\eqref{eq:rho_T} (along with Eqs.~\ref{eq:def_p}-\ref{eq:def_T}) is formally equivalent
to the hydrostatic equation for an ideal gas \cite{binney2008}.
This analogy makes sense here.
Indeed, in the mean-field limit,
the pairwise interactions between particles are neglected
--- just like we neglected any collision operator in the right-hand side of Eq.~\eqref{eq:boltzmann}.
As a result, in this limit, the only contribution to the system's internal energy
stems from the particles' disordered motion measured by the temperature.
In that view, the specific internal energy, $u$,
follows from the familiar formula of the specific kinetic energy, ${ \half v^{2} }$,
i.e., one has
\begin{equation}
u=\half T.
\label{eq:uhalfT}
\end{equation}
In 1D, an ideal gas is a system complying exactly with Eqs.~\eqref{eq:def_p} and~\eqref{eq:uhalfT}.

In a 1D self-gravitating system, the potential $\phi$
follows from Poisson equation, namely
\begin{equation}
\phi''= 2 \mG \rho
\quad \text{or equivalently} \quad
g' = - 2 \mG \rho .
\label{eq:poisson}
\end{equation}
with ${ \mG \!>\! 0 }$ the gravitational constant.
Although gravity is driving the evolution,
this does not invalidate our assumption about the lack of collisions.
Indeed, since 1D gravity is a long-range force~\cite{campa2014},
particles dominantly interact with the gravitational background they create
(via Eq.~\ref{eq:poisson})
rather than through pairwise deflections.
Because the present system is self-gravitating,
we can use Poisson equation (Eq.~\ref{eq:poisson}),
to rewrite the hydrostatic relation from Eq.~\eqref{eq:rho_T} as
\begin{equation}
\frac{2 \mG p'}{-g'} = - g g'.
\label{eq:hydro}
\end{equation}
Using Eqs.~\eqref{eq:def_p}-\eqref{eq:uhalfT}, it can be rephrased as
\begin{equation}
\left(2ug'\right)' = gg',
\label{eq:link_ugp}
\end{equation}
Proceeding further, this can be integrated
to give a (local) relation between the internal energy
and the gravitational potential, namely
\begin{equation}
u=\frac{1}{2g'}\left(K+\frac{g^2}{2}\right),
\label{eq:u_phi}
\end{equation}
with $K$ a constant.\footnote{In practice, $K$ is not guaranteed to remain constant once microscopic effects (collisions) are taken into account, i.e., when higher moments are also included.} 
Equation~\eqref{eq:u_phi} is an important relation.
It states that for a 1D self-gravitating system in hydrostatic equilibrium,
the internal energy field, ${ u(x) }$,
is a function of the local gravitational acceleration field, ${g(x)}$,
and its local derivative, ${ g'(x) }$.

Following the previous arguments,
we may treat a 1D self-gravitating system as an ideal gas
embedded within its own self-generated gravitational field.
We now push further this thermodynamic viewpoint.
We assume that we may describe the instantaneous state of the system
using only two moments of the distribution,
namely the density, ${ \rho (x) }$ (Eq.~\ref{eq:def_rho}),
and the temperature, via ${ u (x) }$ (Eq.~\ref{eq:uhalfT}).
Naturally, these fields contain far less information about the system
than the full phase space distribution.
Fortunately, they are still sufficient to explore non-trivial phenomena.

We start with the first law of thermodynamics expressed with quantities per unit mass.
It reads
\begin{equation}
\rd s =\frac{1}{T}\rd u+\frac{p}{T} \, \rd \bigg( \frac{1}{\rho} \bigg) = \frac{1}{T} \rd u - \frac{p}{T\rho^2} \rd \rho ,
\label{eq:2nd_law_init}
\end{equation}
where $s$ is the specific entropy. 
In order to account more easily for self-gravity,
we may recast Eq.~\eqref{eq:2nd_law_init} using Poisson equation (Eq.~\ref{eq:poisson}).
More precisely, we have
\begin{equation}
\rd s = \frac{1}{T} \, \rd u + \frac{2 \mG p}{Tg'^2} \, \rd g' .
\end{equation}
From these differential relations, we can readily derive
\begin{subequations}
\begin{align}
\left( \frac{\p s}{\p u}\right)_{g'} {} & = \frac{1}{T} ,
\label{eq:partial1}
\\
\left( \frac{\p s}{\p g'}\right)_{u} {} & = \frac{2 \mG p}{Tg'^2} .
\label{eq:partial2}
\end{align}
\label{eq:partial_1and2}\end{subequations}
In 1D, the specific entropy of the ideal gas is generically~\cite{Blundell2010}
\begin{equation}
s = \half \ln u - \ln \rho. 
\label{eq:s_init}
\end{equation}
Using Poisson equation (Eq.~\ref{eq:poisson}),
the specific entropy can be expressed
as a function only of the internal energy and gravitational field,
namely
\begin{equation}
s = \half \ln u - \ln \left(-\frac{g'}{2 \mG }\right).
\label{eq:s}
\end{equation}
One can check that injecting Eq.~\eqref{eq:s_init} into Eq.~\eqref{eq:partial_1and2}
recovers, exactly, both Eqs.~\eqref{eq:def_p} and~\eqref{eq:uhalfT}
that define the ideal gas.

With all this, we are now fully equipped to explore the equilibrium states of a 1D self-gravitating system.
In particular, we will investigate the extrema of the system's total entropy.

\subsection{Thermodynamic variations}
\label{sec:entropy_var}

The thermodynamic entropy\footnote{Clarifying its precie relation to the more general Gibbs entropy ${ S \!=\! \! \int \! \rd x \rd v f \ln [f] }$ will be investigated in a future work.} of a self-gravitating system
generically reads
\begin{equation}
S = \!\!\int\!\! \rd x \, \rho (x) \, s [g' (x) , u (x)] ,
\label{eq:def_S_init}
\end{equation}
where we dropped the boundaries of the integral for convenience,
and the function $s$ was introduced in Eq.~\eqref{eq:s}.
Since the system is fully self-gravitating,
it obeys Poisson equation (Eq.~\ref{eq:poisson}).
This allows us to replace the density, $\rho$, with the acceleration, $g'$.
As a result, the entropy then becomes a
functional of the whole gravitational and energy fields,
that we respectively denote with ${ \bg \!=\! \{ g(x) \}_{x} }$\
and similarly ${ \bu \!=\! \{ u (x) \}_{x}}$.
Equation~\ref{eq:def_S_init} becomes
\begin{equation}
S [ \mathbf{g} , \mathbf{u}] = - \frac{1}{2 \mG} \!\! \int \!\! \rd x \, g'(x) \, s[g'(x) , u(x)] .
\label{eq:def_generic_S}
\end{equation}

Following a similar approach,
we can introduce the energy functional.
As detailed in Appendix~\ref{app:energy_1D},
for a generic 1D self-gravitating system it reads
\begin{equation}
E [\bg , \bu] = - \frac{1}{2 \mG} \!\! \int \!\! \rd x \, \bigg[ u (x) g' (x) + \half [g(x)]^2 \bigg],
\label{eq:energy}
\end{equation}
where the first term is the internal ("heat") energy and the second is the self-gravitating binding energy.
Just like the entropy (Eq.~\ref{eq:def_generic_S}),
the energy is a functional of the gravitational and internal energy profiles.

Now that we armed with these two functionals,
we may perform thermodynamic variations of them
by applying perturbations in the fields $\bg$ and $\bu$.
More precisely, we consider a change ${ \bg \!\to\! \bg \!+\! \eps \bD \bg }$
and ${ \bu \!\to\! \bu \!+\! \eps \bD \bu }$,
with ${ (\bD \bg , \bD \bu) }$ some prescribed perturbation fields,
and $\eps$ a small dimensionless parameter.
Importantly, for now, we assume that the respective perturbations,
${ \bD \bg }$ and ${ \bD \bu }$, can be chosen independently
from one another.

In the limit of small perturbations,
i.e.\ in the limit ${ \eps \!\ll\! 1 }$,
the variations of the entropy functional is generically given by
\begin{equation}
S [\bg \!+\! \eps \bD \bg , \bu \!+\! \eps \bD \bu ] \simeq S + \eps \, \Delta S + \half \, \eps^{2} \, \Delta^{2} S + ... ,
\label{eq:variations_S}
\end{equation}
where we shortened the notations in right hand side,
by writing ${ S \!=\! S [\bg , \bu] }$,
${ \Delta S \!=\! \Delta S [\bg , \bu ; \bD \bg , \bD \bu] }$,
and similarly for ${ \Delta^{2} S }$.
The same perturbative expansion also holds for the total energy.
In Eq.~\eqref{eq:variations_S}, the leading-order variation is generically given by
\begin{equation}
\Delta S = \!\! \int \!\! \rd x \, \bigg[ \frac{\delta S}{\delta g (x)} \, \Delta g (x) + \frac{\delta S}{\delta u (x)} \, \Delta u(x) \bigg] ,
\label{eq:first_DeltaS}
\end{equation}
In that expression, ${ \delta / \delta g(x) }$ stands for the functional derivative,
which satisfies the usual fundamental identities
\begin{align}
{} & \frac{\delta g (y)}{\delta g(x)} = \deltaD (x \!-\! y) ;
\quad
{} && \frac{\delta g(y)}{\delta u(x)} = 0 ,
\nonumber
\\
{} & \frac{\delta u(y)}{\delta g(x)} = 0 ;
\quad
{} && \frac{\delta u(y)}{\delta u(x)} = \deltaD (x \!-\! y) .
\label{eq:fundamental_identity}
\end{align}
with $\deltaD$ the Dirac delta.
The second-order variation in Eq.~\eqref{eq:variations_S} reads
\begin{align}
\Delta^{2} S = \!\! \int \!\! \rd x \rd y \, \bigg[ {} & \frac{\delta^{2} S}{\delta g (x) \delta g (y)} \, \Delta g (x) \, \Delta g (y) + \frac{\delta^{2} S}{\delta u(x) \delta u(y)} \, \Delta u(x) \, \Delta u(y)
\nonumber
\\
+ {} & \, 2 \, \frac{\delta^{2} S}{\delta g(x) \delta u(y)} \, \Delta g(x) \, \Delta u (y) \bigg] .
\label{eq:second_DeltaS}
\end{align}
Similar expressions also hold for the energy functional, ${ E \!=\! E [\bg , \bu] }$.

Physically, from Eq.~\eqref{eq:variations_S},
equilibria correspond to extrema of the entropy,
i.e.\ states ${ (\bg , \bu) }$ for which ${ \Delta S \!=\! 0 }$,
whatever the imposed perturbations ${ (\bD \bg , \bD \bu) }$.
Such equilibria are then thermodynamically stable
if they correspond to maxima of the entropy,
i.e.\ if one has ${ \Delta^{2} S \!<\! 0 }$,
whatever ${ (\bD \bg , \bD \bu) }$.

We are now set to use the present formalism in two regimes.
First, we will recover the well-known results
that systems in hydrostatic equilibrium with a uniform temperature profile
are entropy maxima (Section~\ref{sec:Thermo_Eq}).
This is no surprise since these are known to be thermodynamic equilibria.
Then, we will proceed with our main result:
we will show that 1D homogeneous systems
in hydrostatic equilibrium, albeit they are not thermodynamic equilibria,
are also entropy maxima,
when submitted to perturbations that preserve the hydrostatic equilibrium (Section~\ref{sec:tb}).
This is what we refer to as thermodynamic blocking.

\subsection{Thermodynamic equilibrium}
\label{sec:Thermo_Eq}

We consider a generic 1D self-gravitating system,
characterised by some profiles ${ (\bg , \bu) }$.
Our goal is to characterise the properties of this system's entropy variations.

\textit{First-order variations}. 
Following Eq.~\eqref{eq:first_DeltaS},
to compute ${ \Delta S }$,
we need to compute two functional derivatives.
As detailed in Appendix~\ref{app:1D_therm_details_1der},
we find that the leading variation of the entropy is given by
\begin{equation}
\Delta S = \frac{1}{2 \mG} \!\! \int \!\! \rd x \, \bigg[ - \frac{g' (x)}{2u (x)} \, \Delta u (x) + \bigg\{ \frac{u' (x)}{2u (x)} - \frac{g''(x)}{g'(x)} \bigg\} \, \Delta g (x) \bigg] .
 \label{eq:delta_S_calculated}
\end{equation}
At this stage, we note that the vanishing of ${ \Delta S }$
at thermodynamical equilibrium is not yet obvious.
To make progress, we may perform a similar calculation for ${ \Delta E }$.
As detailed in Appendix~\ref{app:1D_therm_details_1der},
we start from Eq.~\eqref{eq:energy} to find
\begin{equation}
\Delta E =\frac{1}{2 \mG} \!\! \int \!\! \rd x \, \bigg[ - g' (x) \, \Delta u (x) + [u' (x) - g (x)] \, \Delta g (x) \bigg] .
\label{eq:DeltaE_almost}
\end{equation}

Let us now place ourselves at thermodynamical equilibrium.
As such, we consider states ${ (\bg , \bu) }$ such that
\begin{equation}
\begin{cases}
\displaystyle u = u_{0} = \cst & \text{(Constant internal energy profile)} ,
\\
\displaystyle (2 u g^{\prime})^{\prime} = g g^{\prime} & \text{(Hydrostatic equilibrium --- Eq.~\ref{eq:link_ugp})} .
\end{cases}
\label{eq:def_rT}
\end{equation}
For the variation of the energy,
we use ${ u^{\prime} \!=\! 0 }$ in Eq.~\eqref{eq:DeltaE_almost},
and find
\begin{equation}
\Delta E \big|_{\rT} = -\frac{1}{2 \mG} \!\!\int\!\! \rd x \, \bigg[ g' (x) \Delta u (x) + g (x) \Delta g (x) \bigg] ,
\label{eq:DeltaE_rTE}
\end{equation}
where the subscript "${ |_{\rT} }$" emphasises that we are computing
variations around a thermodynamical equilibrium.
For the variation of the entropy,
using Eq.~\eqref{eq:def_rT},
we may impose ${ 2 u_{0} g^{\prime\prime} \!=\! g g^{\prime} }$
in Eq.~\eqref{eq:delta_S_calculated}.
We obtain
\begin{equation}
\Delta S \big|_{\rT} = -\frac{1}{4 \mG u_0} \!\!\int\!\! \rd x \, \bigg[ g' (x) \Delta u (x) + g (x) \Delta g (x) \bigg].
\label{eq:DeltaS_rTE}
\end{equation}

Now, comparing Eqs.~\eqref{eq:DeltaE_rTE} and~\eqref{eq:DeltaS_rTE},
we find that, at thermodynamical equilibrium,
the first-order variations in the energy and entropy
are simply linked via\footnote{Using Eq.~\eqref{eq:uhalfT}, this formula translates into the well-known ${ \rd E \!=\!T \rd S }$.} 
\begin{equation}
\Delta S \big|_{\rT} = \frac{1}{2u_{0}} \, \Delta E \big|_{\rT} .
\label{eq:equality_dE_dS}
\end{equation}

In practice, here we are focusing on internally-generated perturbations.
Phrased differently, the only variations ${ (\bD \bg , \bD \bu) }$ that are physically
allowed are the ones that conserve the total energy.
As a consequence, one must have ${ \Delta E \big|_{\rT} \!=\! 0 }$.
From Eq.~\eqref{eq:equality_dE_dS},
we immediately conclude that, at thermodynamical equilibrium,
one has
\begin{equation}
\Delta S \big|_{\rT} = 0 .
\label{eq:vanishing_dS_TE}
\end{equation}
This is a reassuring result.
It shows that, within the present fluid formalism,
thermodynamical equilibria are, indeed, entropy extrema. Note, however, that Eq.~\eqref{eq:vanishing_dS_TE} is only of first-order accuracy.
Examining the second-order variation is crucial
to determine the sign of the entropy variation.

\textit{Second-order variations}. 
Let us now push the present calculation further
and compute the second-order variation of the entropy and the energy,
at thermodynamical equilibrium.
We follow the definition from Eq.~\eqref{eq:second_DeltaS}.
As detailed in Appendix~\ref{app:1D_therm_details_2der},
we find
\begin{align}
\Delta^{2} S = \frac{1}{2 \mG} \!\! \int \!\! \rd x \, \bigg[ \frac{[\Delta g' (x)]^{2}}{g'(x)} \!-\! \frac{\Delta g'(x) \, \Delta u (x)}{u (x)} \!+\! \tfrac{1}{2} \frac{g'(x) \, [ \Delta u (x) ]^{2}}{[u(x)]^{2}} \bigg] .
\label{eq:D2_S_poisson_square}
\end{align}
where we introduced the notation ${ \Delta g' (x) \!=\! \rd \Delta g (x) / \rd x }$.
At this stage, the sign of ${ \Delta^{2} S }$ is not obvious. To make progress, we account for the second-order variations in the energy, which are calculated in Appendix~\ref{app:1D_therm_details_2der}.
We find
\begin{equation}
\Delta^{2} E = - \frac{1}{2 \mG} \!\! \int \!\! \rd x \, \bigg[ \big[ \Delta g (x) \big]^{2} \!+\! 2 \Delta g' (x) \Delta u(x) \bigg] .
\label{eq:D2_E_1D}
\end{equation}

We now demand energy conservation
up to second-order accuracy,
i.e. we impose
\begin{equation}
\Delta E \!+\! \tfrac{1}{2}\Delta ^2 E \!=\! 0 . 
\label{eq:energy_cons_1st_2nd}
\end{equation}
Using Eqs.~\eqref{eq:delta_S_calculated} and \eqref{eq:D2_S_poisson_square} to calculate the total entropy variation 
\begin{equation}
\Delta S_{\rtot} \!=\! \Delta S \!+\! \tfrac{1}{2}\Delta ^2S ,
\label{eq:Stot}
\end{equation}
placing ourselves into thermodynamic equilibrium determined by Eq.~\eqref{eq:def_rT}, one finally arrives at
\begin{equation}
\Delta S_{\rtot} \big|_{\rT}=\frac{1}{4 \mG} \!\! \int \!\! \rd x \, \bigg[ \frac{g'(x)[\Delta u(x)]^2}{2u_0^2}+\frac{[\Delta g(x)]^2}{2u_0} + \frac{[\Delta g'(x)]^2}{g'(x)} \bigg].
\label{eq:deltaStot}
\end{equation}
As expected, there is no first-order variation in the total entropy change. Regarding its sign, we observe that the first term in the right-hand side of Eq.~\eqref{eq:deltaStot} is negative, since ${ g' \!\propto\! - \rho \!<\! 0}$ (Eq.~\ref{eq:poisson}). Determining the sign of the remaining two terms is Eq.~\eqref{eq:deltaStot} is much more challenging.
We will not pursue this calculation here,
and refer to Appendix~{B} in~\cite{padmanabhan1989}
for a similar calculation in the 3D case.
Nonetheless, let us give two heuristic arguments to support the (physically-motivated) fact that the sum of the second and third terms in Eq.~\eqref{eq:deltaStot} is negative.
First, in 1D, the thermodynamic equilibrium is
${ g' \!\propto\! -\cosh^{-2}(x) }$~\citep[see, e.g.\@,][]{roule2022}.
As a result, the third term of Eq.~\eqref{eq:poisson}, which is negative, is expected to dominate over the second for ${ x \!\gg\! 1 }$.
Second, as pointed out by~\cite{padmanabhan1989},
if one confines the system in a sufficiently small (1D) box,
the present system can be approximated as an homogeneous ideal gas with negligible self-gravity. This is known to be of maximal entropy.

As a closing remark, we note that the present calculation does not guarantee that the thermodynamic equilibria physically exist.
Fortunately, the 1D thermodynamical equilibria genuinely exist~\citep[see, e.g.\@,][]{Miller+2023},
contrary to 3D~\citep{chavanis2002}.

\subsection{Thermodynamic blocking}
\label{sec:tb}

We now delve into the main result of this work,
namely the existence in 1D self-gravitating systems
of another way in which entropy may set an extremum.
We note that in any self-gravitating system,
the finite number of particles, $N$, unavoidably sets
some (weak) level of collisionality in the system.
As such, Poisson fluctuations always tend to drive relaxation.
In practice, given that ${ N \!\gg\! 1 }$, this evolution is slow
so that the systems evolves, adiabatically\footnote{In the present context,
the term "adiabatic" simply refers to slow processes (like, for example, adiabatic invariants in mechanics), and not ones with zero entropy change.},
through a series
of hydrostatic equilibria,
i.e.\ through a series of states, ${ (\bg , \bu) }$,
that instantaneously comply with Eq.~\eqref{eq:u_phi}.
Phrased differently, during such an evolution,
the instantaneous state of the system can be described solely with $\bg$,
leveraging the constraint from Eq.~\eqref{eq:u_phi}.
Along the same line of thought,
physically allowed fluctuations,
because they must comply with the hydrostatic equilibrium,
can be fully characterised by ${ \bD \bg }$.

Let us now write the entropy functional
for such a system that evolves only along
hydrostatic equilibria.
Injecting Eq.~\eqref{eq:u_phi} into Eq.~\eqref{eq:s},
we can rewrite the specific entropy
as a sole function of the local gravitational fields (and its derivatives).
It reads
\begin{equation}
s_{\rA} = \half \ln \big( \!-K - \half g^{2} \big) - \tfrac{3}{2} \ln(-g') ,
\label{eq:s_A}
\end{equation}
where specical care must be taken with the signs
in the arguments of the $\ln$ --- since ${ g' \!<\! 0 }$
owing to Poisson equation (Eq.~\ref{eq:poisson}) ---
and we dropped a constant term for convenience.\footnote{These omitted terms give zero contribution to the entropy change due to mass conservation.}
In Eq.~\eqref{eq:s_A}, the subscript "$\rA$" emphasises
that we are considering systems submitted to adiabatic perturbations.

Following Eq.~\eqref{eq:def_generic_S},
the total entropy becomes now a sole function
of the gravitational field, $\bg$.
It reads
\begin{equation}
S_{\rA} [\bg] = - \frac{1}{2 \mG} \!\! \int \!\! \rd x \, g'(x) \, s_{\rA} [g(x) , g'(x)] .
\label{eq:func_S_A}
\end{equation}
Similarly, injecting Eq.~\eqref{eq:u_phi} into Eq.~\eqref{eq:energy},
we can rewrite the total energy as a functional
of only the gravitational field, $\bg$. It reads
\begin{equation}
E_{\rA} [\bg] = - \frac{3}{8 \mG} \!\! \int \!\! \rd x \, [g (x)]^{2} ,
\label{eq:def_EA}
\end{equation}
where we dropped the constant term in $K$.

When submitted to some perturbation ${ \bD \bg }$,
similarly to Eq.~\eqref{eq:variations_S},
the variations of the entropy are given by
\begin{equation}
S_{\rA} [\bg \!+\! \eps \bD \bg] \simeq S_{\rA} + \eps \, \Delta S_{\rA} + \half \, \eps^{2} \, \Delta^{2} S_{\rA} + ... ,
\label{eq:variations_S_A}
\end{equation}
with the shortened notations, ${ S_{\rA} \!=\! S_{\rA} [\bg] }$,
${ \Delta S_{\rA} \!=\! \Delta S_{\rA} [\bg ; \bD \bg] }$,
and similarly for ${ \Delta^{2} S }$.
Following Eqs.~\eqref{eq:first_DeltaS} and~\eqref{eq:second_DeltaS},
the variations of the entropy under adiabatic perturbations read
\begin{subequations}
\begin{align}
\Delta S_{\rA} {} & = \!\! \int \!\! \rd x \, \frac{\delta S_{\rA}}{\delta g(x)} \, \Delta g (x)
\label{eq:first_DeltaS_A}
\\
\Delta^{2} S_{\rA} {} & = \!\! \int \!\! \rd x \rd y \, \frac{\delta^{2} S_{\rA}}{\delta g(x) \delta g(y)} \, \Delta g(x) \, \Delta g(y) .
\label{eq:second_DeltaS_A}
\end{align}
\label{eq:DeltaS_A}\end{subequations}
Naturally, similar relations also hold for the variations of the total energy.
We are now set to compute explicitly the first- and second-order
variations of the present "adiabatic" entropy and energy.

These calculations are detailed in Appendix~\ref{app:1D_blck_details}.
We find that the variations of the total entropy are given by
\begin{subequations}
\begin{align}
\Delta S_{\rA} {} & = - \frac{3}{4 \mG} \!\! \int \!\! \rd x \, \frac{g''(x)}{g'(x)} \, \Delta g (x) ,
\label{eq:exp_DeltaS_A_1st}
\\
\Delta^{2} S_{\rA} {} & = \frac{3}{4 \mG} \!\! \int \!\! \rd x \, \frac{[\Delta g' (x)]^{2}}{g'(x)} .
\label{eq:exp_DeltaS_A_2nd}
\end{align}
\label{eq:exp_DeltaS_A}\end{subequations}
while those of the total energy read
\begin{subequations}
\begin{align}
\Delta E_{\rA} {} & = - \frac{3}{4 \mG} \!\! \int \!\! \rd x \, g(x) \, \Delta g (x) ,
\label{eq:exp_DeltaE_A_1st}
\\
\Delta^{2} E_{\rA} {} & = - \frac{3}{4 \mG} \!\! \int \!\! \rd x \, [\Delta g (x)]^{2} .
\label{eq:exp_DeltaE_A_2nd}
\end{align}
\label{eq:exp_DeltaE_A}\end{subequations}

As a first sanity check,
let us consider the case of a thermodynamical equilibrium
undergoing adiabatic perturbations.
Imposing the constraints from Eq.~\eqref{eq:def_rT},
we find from Eqs.~\eqref{eq:exp_DeltaS_A_1st} and~\eqref{eq:exp_DeltaE_A_1st}
that, at first order, the variations in the total entropy and total energy are connected via
\begin{equation}
\Delta S_{\rA} \big|_{\rT} = \frac{1}{2 u_{0}} \, \Delta E_{\rA} \big|_{\rT} .
\label{eq:link_DeltaA_T}
\end{equation}

Let us now consider the second-order variation
of the entropy.
Imposing energy conservation up to second-order
(as in Eq.~\ref{eq:energy_cons_1st_2nd}),
we find that the total entropy variation
at second order,
namely
\begin{equation}
\Delta S_{\rA , \rtot} = \Delta S_{\rA} + \tfrac{1}{2} \Delta^{2} S_{\rA}
\label{eq:def_SAtot}
\end{equation}
is
\begin{equation}
\Delta S_{\rA , \rtot} \big|_{\rT} = \frac{3}{8 \mG} \!\! \int \!\! \rd x \, \bigg[ \frac{[\Delta g' (x)]^{2}}{g'(x)} + \frac{[\Delta g(x)]^{2}}{2 u_{0}} \bigg] .
\label{eq:DStot_therm_adiab}
\end{equation}
As could have been expected,
Eq.~\eqref{eq:DStot_therm_adiab} bears deep similarities
with Eq.~\eqref{eq:deltaStot} obtained when considering arbitrary perturbations.
In particular, up to a global (positive) prefactor,
both expressions match if one assumes ${ \Delta u (x) \!=\! 0 }$
in Eq.~\eqref{eq:deltaStot}.
At hydrostatic equilibrium,
one has ${ g' \!\propto\! - \rho \!<\! 0 }$ (Eq.~\ref{eq:poisson}),
so that the first term in Eq.~\eqref{eq:DStot_therm_adiab}
is clearly negative.
As previously, the overall sign of the full expression remains not obvious.
We do not pursue this calculation further
and, once again, refer to Appendix~{B} in~\cite{padmanabhan1989}.

Now comes the main result of our present work.
Indeed, the crucial point is to note these are not the only entropy extrema
under such perturbations. Indeed, let us now consider 1D
homogeneous self-gravitating systems in hydrostatic equilibrium.
As such, we consider systems characterised by some $\bg$ such that
\begin{equation}
\begin{cases}
\displaystyle \rho = \cst & \text{(Homogeneous density)} ,
\\
\displaystyle (2 u g^{\prime})^{\prime} = g g^{\prime} & \text{(Hydrostatic equilibrium --- Eq.~\ref{eq:link_ugp})} .
\end{cases}
\label{eq:def_rH}
\end{equation}
Because such systems are homogeneous,
one has ${ \rho^{\prime} \!=\! 0 }$,
which when plugged in Poisson equation (Eq.~\ref{eq:poisson})
gives ${ g'' \!=\! 0 }$.
As a result, we find from Eq.~\eqref{eq:exp_DeltaS_A_1st}
that the first-order variation of the entropy vanishes for homogeneous
states. Given that such systems also comply with Poisson equation (Eq.~\ref{eq:poisson}),
we can generically rewrite the second-order variation from Eq.~\eqref{eq:exp_DeltaS_A_2nd} as
\begin{equation}
\Delta^{2} S_{\rA} = - \frac{3}{8 \mG^{2}} \!\! \int \!\! \rd x \, \frac{[\Delta g'(x)]^{2}}{\rho (x)} .
\label{eq:DSAtot_rewrite}
\end{equation}
This is always negative, since ${ \rho (x) \!<\! 0 }$.
As a result, we therefore have
\begin{equation}
\Delta S_{\rA} \big|_{\rH} = 0 ;
\quad
\Delta^{2} S_{\rA} \big|_{\rH} < 0 ,
\label{eq:conclusion_DeltaA_H}
\end{equation}
where the subscript "$|_{\rH}$" emphasizes
that we are computing variations around an homogeneous
hydrostatic equilibrium.
In Appendix~\ref{app:barycenter},
we provide an alternative derivation of the thermodynamic blocking,
by leveraging the conservation of the barycentre.

This is a main result of this paper.
It states that homogeneous 1D systems,
in hydrostatic equilibrium,
are entropy maxima
with respect to adiabatic perturbations,
i.e.\ perturbations that, themselves, maintain the hydrostatic equilibrium.
This is a non-trivial result since homogeneous distributions,
as in Eq.~\eqref{eq:link_ugp} are not global thermodynamical equilibrium,
since their profile of internal energy is not constant.
Phrased differently, in a 1D homogeneous system,
heat does not flow from a hotter region to a colder one (under the assumptions we made). 
This is what we refer to as a "thermodynamic blocking".

As mentioned earlier, during the relaxation of a self-gravitating system,
Poisson fluctuations drive the variations in the internal energy
and the gravitational acceleration.
As such, the perturbation ${ \bD \bg }$
is typically expected to scale like ${ N^{-1/2} }$.
In turn, this gives a hint about the scale of accuracy of the present
thermodynamic blocking.
Importantly, in deriving Eq.~\eqref{eq:exp_DeltaS_A_2nd}, we did not need to assume homogeneity.
That is, hydrostatic entropy extrema are always maxima in the adiabatic regime,
i.e.\ one always have ${ \Delta^{2} S_{\rA} \!<\! 0 }$, following Eq.~\eqref{eq:exp_DeltaS_A_2nd}.
This is a limitation of the formalism,
since an homogeneous system must eventually reach another maximum,
namely the thermodynamic equilibrium.
To do so, higher-order microscopic effects must be present.
These are accounted for by higher moments of the distribution function,
starting from Eq.~\eqref{eq:boltzmann}.

Finally, in Appendix~\ref{app:Jeans}, we provide another illustration
of the present thermodynamic formalism
by recovering the properties of the Jeans instability
in 1D self-gravitating systems.

\section{3D self-gravitating spheres}
\label{sec:3d}

We now apply all the previous ideas to 3D self-gravitating spheres.
Given their astrophysical importance, homogeneous 3D spheres
have been extensively studied, both numerically and analytically.
In particular, \cite{Read+2006} showed how the usual formula
for dynamical friction~\cite{Chandrasekhar1943,Tremaine+1984}
fails in homogeneous spheres. This leads to super-dynamical friction~\citep{Zelnikov+2016},
along with core stalling~\citep{Kaur+2022} and dynamical buoyancy~\citep{Banik+2022}.
This was also visited numerically in~\citep{Sellwood2015},
which showed that indeed, the self-consistent relaxation
of a homogeneous sphere is significantly delayed compared
to other non-uniform density profiles.
In the present section, building upon the previous thermodynamic formalism,
we show that homogeneous self-gravitating spheres
in hydrostatic equilibrium are also thermodynamically blocked
when submitted to spherically symmetric adiabatic perturbations.
Given that the mathematical details are essentially the same in 1D and 3D,
we will restrict ourselves to the main physical results in the coming sections.

\subsection{Thermodynamic variations}
\label{sec:variations_3D}

Following the same arguments as in Section~\ref{sec:kinetic_formalism},
one can readily recast the kinetic equations of a 3D system into a thermodynamic formalism.
More precisely, in 3D, the ideal gas laws read~\cite{Blundell2010}
\begin{subequations}
\begin{align}
u = \tfrac{3}{2} T
\label{ideal_gas_3D_u}
\\
p = \rho T
\label{eq:ideal_gas_3D_rho}
\end{align}
\label{eq:ideal_gas_3D}\end{subequations}
And, similarly to Eq.~\eqref{eq:rho_T},
a system will be in hydrostatic equilibrium
if one has
\begin{equation}
p' = \rho g .
\label{eq:hydrostatic_3D}
\end{equation}
Following the same approach as in Eq.~\eqref{eq:s_init},
the specific entropy for the 3D ideal gas
is generically
\begin{equation}
s=\tfrac{3}{2}\ln u - \ln \rho.
\label{eq:def_s_3D}
\end{equation}
Using the expression from Eq.~\eqref{eq:def_s_3D}, one can check that computing the differentials in Eq.~\eqref{eq:2nd_law_init}
 recovers, exactly,
the ideal gas laws in 3D, as in Eq.~\eqref{eq:ideal_gas_3D}.
In 3D, self-gravitating systems
can have a negative heat capacity,
which leads to the so-called
"gravo-thermal catastrophe"~\cite{LyndenBell+1968}.

In practice, we assume that the system is spherically symmetric,
and introduce its enclosed mass profile, ${ m (r) }$,
with ${ r \!=\! |\br| }$ the radius.
The enclosed mass is connected to the density via
\begin{equation}
\rho (r) = \frac{m' (r)}{4 \pi r^{2}} .
\label{eq:link_rho_m}
\end{equation}
In addition, because of spherical symmetry,
Newton's shell theorem connects the gravitational acceleration
to the enclosed mass via\footnote{Note that we denoted the gravitational constant with the same symbol ($\mG$) in both 1D and 3D.}
\begin{equation}
g(r) = - \frac{\mG m (r)}{r^{2}} .
\label{eq:Newton_shell_3D}
\end{equation}
Following the same approach as in Eq.~\eqref{eq:def_S_init},
the total entropy is generically
\begin{equation}
S = \!\! \int \!\! \rd \br \, \rho (\br) \, s [u(\br) , \rho(\br)].
\label{eq:def_S_3D_start}
\end{equation}
At this stage, we recall that 3D systems have well-defined equilibria only if they are enclosed in a finite volume.\footnote{This is not a sufficient condition: temperature must also be above a certain threshold; see \cite{binney2008} for further details.} As a consequence, the integral here and below are meant between $0$ and some finite radius. Leveraging spherically symmetry and Eq.~\eqref{eq:link_rho_m},
we can rewrite Eq.~\eqref{eq:def_S_3D_start} as
\begin{equation}
S [\bu , \bm] = \!\! \int \!\! \rd r \, m' (r) \, \big[ \tfrac{3}{2} \ln [u (r)] - \ln [\rho (r)] \big] .
\label{eq:def_S_3D}
\end{equation}
Here, we view the total entropy as a functional
of both the internal energy and enclosed mass profiles, $\bu$ and $\bm$.
In particular, following Eq.~\eqref{eq:link_rho_m},
we have ${ \rho (r) \!=\! \rho[\bm] (r) }$.
Following a similar approach, we may introduce the energy functional.
As detailed in Appendix~\ref{app:energy_3D},
for a 3D self-gravitating system,
it can be expressed as a function of ${ (\bu , \bm) }$ via 
\begin{equation}
E[\bu , \bm]= \!\! \int \!\! \rd r \, \bigg[ m'(r)u(r) - \frac{\mG m(r) m'(r)}{r} \bigg],
\label{eq:def_E_3D}
\end{equation}

Just like we did in 1D, let us now assume that the system is submitted
to some perturbations ${ \bu \!\to\! \bu \!+\! \eps \bD \bu }$
and ${ \bm \!\to\! \bm \!+\! \eps \bD \bm }$.
Importantly, we assume that these perturbations are spherically symmetric,
i.e.\ ${ \Delta u (\br) \!=\! \Delta u (r) }$
and similarly for ${ \Delta m }$.
We stress that this is a particularly stringent assumption,
which makes the upcoming calculation tractable.
Naturally, it would be interesting to go beyond this assumption
and consider perturbations with arbitrary symmetries.
This will be the topic of a future work.
Following Eq.~\eqref{eq:first_DeltaS},
the first-order variation of the total entropy
under such generic spherically symmetric perturbations reads
\begin{equation}
\Delta S = \!\! \int \!\! \rd r \, \bigg[ \frac{\delta S}{\delta u (r)} \, \Delta u (r) + \frac{\delta S}{\delta m (r)} \, \Delta m (r) \bigg] .
\label{eq:DeltaS_3D}
\end{equation}
Similarly, following Eq.~\eqref{eq:second_DeltaS},
the second-order variation of the total entropy reads
\begin{align}
\Delta^{2} S = \!\! \int \!\! \rd r \rd \tr \bigg[ {} & \frac{\delta^{2} S}{\delta u (r) \delta u (\tr)} \, \Delta u (r) \, \Delta u(\tr) + \frac{\delta^{2} S}{\delta m (r) \delta m(\tr)} \, \Delta m (r) \, \Delta m(\tr) 
\nonumber
\\
+ \, {} & 2 \frac{\delta^{2} S}{\delta u (r) \delta m(\tr)} \, \Delta u(r) \, \Delta m (\tr) \bigg] .
\label{eq:second_DeltaS_3D}
\end{align}
Naturally, analogous expressions hold for the energy variation.

\subsection{Thermodynamic equilibrium}
\label{sec:3d:TE}

We now follow the same approach as in Section~\ref{sec:Thermo_Eq}
and consider a generic 3D self-gravitating system,
characterised by some profiles ${ (\bu , \bm) }$.
Our goal is to characterise the properties of this system's entropy variations.

\textit{First-order variations}. Following Eq.~\eqref{eq:DeltaS_3D},
to compute ${ \Delta S }$, we need to compute two functional derivatives.
As detailed in Appendix~\ref{app:3D_thrm_details_1der}, the leading variation of the entropy functional
is given by
\begin{equation}
\Delta S = \!\! \int \!\! \rd r \, \bigg[ \tfrac{3}{2} \frac{m' (r)}{u (r)} \, \Delta u (r) + \bigg(  \frac{\rho'(r)}{\rho (r)} - \tfrac{3}{2} \frac{u'(r)}{u(r)} \bigg) \, \Delta m (r) \bigg] .
\label{eq:DeltaS_3D_generic}
\end{equation}
Similarly to the 1D case, it proves fruitful
to compute the first-order variation of the total energy.
As detailed in Appendix~\ref{app:3D_thrm_details_1der}, it reads
\begin{equation}
\Delta E = \!\! \int \!\! \rd r \, \bigg[ m'(r) \, \Delta u(r) - \left(u'(r)+\frac{\mG m (r)}{r^{2}}\right) \, \Delta m (r) \bigg] .
\label{eq:DeltaE_3D_generic}
\end{equation}

We now place ourselves at thermodynamical equilibrium,
i.e.\ we consider states ${ (\bu, \bm) }$ such that
\begin{equation}
\begin{cases}
\displaystyle u = u_{0} = \cst & \text{(Constant internal energy profile)} ,
\\
\displaystyle p' = \rho g & \text{(Hydrostatic equilibrium --- Eq.~\ref{eq:hydrostatic_3D})} .
\end{cases}
\label{eq:def_rT_3D}
\end{equation}
Using Eq.~\eqref{eq:ideal_gas_3D} in conjunction with Eq.~\eqref{eq:hydrostatic_3D},
we find that at thermodynamical equilibrium, one has
\begin{subequations}
\begin{align}
p {} & = \frac{2}{3 u_{0}} \, \rho
\label{eq:link_eqT_3D_ptorho}
\\
\frac{p'}{p} {} & = \frac{\rho'}{\rho} = \frac{3}{2 u_{0}} \, g.
\end{align}
\label{eq:link_eqT_3D}\end{subequations}
Using these relations,
we can rewrite the entropy variation from Eq.~\eqref{eq:DeltaS_3D_generic} as
\begin{equation}
\Delta S \big|_{\rT} = \frac{3}{2 u_{0}} \! \int \!\! \rd r \, \bigg[ m'(r) \, \Delta u (r) + g(r) \, \Delta m (r) \bigg] .
\label{eq:calc_DS_3D_Thr}
\end{equation}
Similarly, we can rewrite the energy variation from Eq.~\eqref{eq:DeltaE_3D_generic} as
\begin{equation}
\Delta E \big|_{\rT} = \!\! \int \!\! \rd r \, \bigg[  m'(r) \, \Delta u (r) + g (r) \, \Delta m (r) \bigg] ,
\label{eq:calc_DE_3D_Thr}
\end{equation}
where we used the relations from Eqs.~\eqref{eq:link_rho_m} and~\eqref{eq:Newton_shell_3D}.
Now, comparing Eqs.~\eqref{eq:calc_DS_3D_Thr} and~\eqref{eq:calc_DE_3D_Thr},
we find therefore that, at thermodynamical equilibrium,
the first-order variations in the entropy and energy are simply linked via
\begin{equation}
\Delta S \big|_{\rT} = \frac{3}{2 u_{0}} \, \Delta E \big|_{\rT} .
\label{eq:equality_DS_DE_3D}
\end{equation}
Naturally, this bears similarities with the relation from Eq.~\eqref{eq:equality_dE_dS}
obtained in the 1D case.
Since we are focusing on internally-generated (spherically symmetric) perturbations, ${ (\bD \bu , \bD \bm) }$,
we must have ${ \Delta E |_{\rT} \!=\! 0 }$ up to first-order accuracy,
to comply with energy conservation.
As a consequence, from Eq.~\eqref{eq:equality_DS_DE_3D},
we find that, at first order,
at thermodynamical equilibrium, one has
\begin{equation}
\Delta S \big|_{\rT} = 0 .
\label{eq:vanish_S_3D_Thr}
\end{equation}
This is good news. We recover that, in 3D,
the thermodynamical equilibria are indeed entropy extrema. Note, however, that in deriving this result we considered spherically symmetric perturbations only. A more general study is left for future work.

\textit{Second-order variations}. 
Following the same machinery as in 1D, we use second-order energy variation in order to determine the overall sign of entropy variation. Appendix~\ref{app:3D_thrm_details_2der} presents the detailed second-order calculations for both entropy and energy variations. Injecting the latter into the former yields the total entropy variation (Eq.~\ref{eq:Stot})
\begin{align}
\Delta S_{\rtot} \big|_{\rT} =\tfrac{1}{2} \!\! \int \!\! \rd r \rd \tr \, \bigg[ {} & \frac{3}{2 u_0}\frac{\mG \deltaD(r \!-\! \tr)}{r^2} \Delta m(r) \Delta m(\tr)
\nonumber
\\
- {} & \frac{3}{2 u_0^2} m'(r) \deltaD (r \!-\! \tr) \Delta u(r) \Delta u(\tr) - \frac{\deltaD' (r \!-\! \tr)}{m'(r)} \Delta m'(r) \Delta m (\tr) \bigg] . 
\label{eq:final_D2S_thrm_3D} 
\end{align}
After additional manipulations, it becomes
\begin{equation}
    \Delta S_{\rtot} \big|_{\rT} = - \tfrac{1}{2} \!\!\int\!\! \rd r \bigg[ \tfrac{3}{2}\frac{m'(r)[\Delta u(r)]^2}{[u(r)]^2} + \frac{[\Delta m'(r)]^2}{m'(r)} - \tfrac{3}{2} \frac{\mG}{u_0} \frac{[\Delta m(r)]^2}{r^2} \bigg].
\label{eq:DeltaStot_normal_therm}
\end{equation}
Analogously to the 1D case, the first term on the right-hand side is negative. Similarly, the second term in Eq.~\eqref{eq:DeltaStot_normal_therm} is also since negative,
since one has ${ m' \!\propto\! \rho \!>\! 0 }$ (Eq.~\ref{eq:link_rho_m}).
Yet, just like in the 1D case,
determining the sign of the sum of the second and third terms
is much more challenging.
This calculation is presented in detail in Appendix~{B} of \cite{padmanabhan1989}.
Therein, it is shown that thermodynamical equilibria, i.e.\ systems in hydrostatic equilibrium with a constant internal energy profile (Eq.~\ref{eq:def_rT_3D}) are entropy maxima.

\subsection{Thermodynamic blocking}
\label{sec:thrm_blck_3D}

Let us now investigate the possibility of thermodynamic blocking in 3D.
Following the same approach as in 1D (Section~\ref{sec:tb}),
let us assume that our system evolves (slowly) through a series
of hydrostatic equilibria,
i.e.\ through a series of states ${ (\bp , \bm) }$
that instantaneously comply with Eq.~\eqref{eq:hydrostatic_3D}.
We use the pair ${ (\bp, \bm) }$ instead of ${ (\bu , \bm ) }$
to track the system's state: this eases the upcoming calculations.
Let us now assume that our system is in hydrostatic equilibrium.
Given the constraint from Eq.~\eqref{eq:hydrostatic_3D},
we can describe the instantaneous state of the system only using $\bm$.
Combining the relations from Eqs.~\eqref{eq:link_rho_m} and~\eqref{eq:Newton_shell_3D},
we can rewrite the hydrostatic equilibrium condition as
\begin{equation}
p' (r) = - \frac{\mG m (r) m'(r)}{4 \pi r^{4}} .
\label{eq:hydro_3D_expl}
\end{equation}
Here, we point out that the functional relation
between $\bp$ and $\bm$ cannot be given in a closed form,
contrary to the 1D case, where we could express $\bu$ explicitly
as a function of $\bg$ (see Eq.~\ref{eq:u_phi}).
This makes the 3D calculation slightly less transparent,
but it does not affect the results to be derived.

Following Eq.~\eqref{eq:def_S_3D}, the total entropy now becomes
a sole function of the enclosed mass profiles, $\bm$. It reads
\begin{equation}
S_{\rA} [\bm] = \!\! \int \!\! \rd r \, m' (r) \, \big[ \tfrac{3}{2} \ln [p (r)] - \tfrac{5}{2} \ln [\rho (r)] \big] ,
\label{eq:def_SA_3D}
\end{equation}
where we stress that both the pressure and density profiles
are to be interpreted as functionals of the enclosed mass profiles,
i.e.\ one has ${ p (r) \!=\! p [\bm] (r) }$ (via Eq.~\ref{eq:hydro_3D_expl})
and ${ \rho (r) \!=\! \rho [\bm] (r) }$ (via Eq.~\ref{eq:link_rho_m}).
Similarly, following Eq.~\eqref{eq:def_EA_3D},
the total energy now reads
\begin{equation}
E_{\rA} [\bm]= \!\! \int \!\! \rd r \, \bigg[ \tfrac{3}{2} 4\pi r^2 p (r) - \frac{\mG m(r) m'(r)}{r} \bigg] ,
\label{eq:def_EA_3D}
\end{equation}
with the same self-consistent constraint to be satisfied by ${ p (r) }$.
As previously, the subscript "$|_{\rA}$" emphasises
that we are considering systems in hydrostatic equilibrium.

Just like we did in 1D, we now assume that the system is submitted
to some (spherically symmetric) adiabatic perturbations, ${ \bm \!\to\! \bm \!+\! \eps \bD \bm }$.
Following Eq.~\eqref{eq:DeltaS_A},
the variations of the total entropy read
\begin{subequations}
\begin{align}
\Delta S_{\rA} {} & = \!\! \int \!\! \rd r \, \frac{\delta S_{\rA}}{\delta m(r)} \, \Delta m (r)
\label{eq:first_DeltaS_A_3D}
\\
\Delta^{2} S_{\rA} {} & = \!\! \int \!\! \rd r \rd \tr \, \frac{\delta^{2} S_{\rA}}{\delta m(t) \delta m (\tr)} \, \Delta m (r) \, \Delta m (\tr) .
\label{eq:second_DeltaS_A_3D}
\end{align}
\label{eq:DeltaS_A_3D}\end{subequations}
As detailed in Appendix~\ref{app:3D_blck_details},
we find that the variations of the total entropy are generically given by
\begin{subequations}
\begin{align}
\Delta S_{\rA} {} & = \tfrac{5}{2} \!\! \int \!\! \rd r \, \frac{\rho' (r)}{\rho (r)} \, \Delta m (r) ,
\label{eq:exp_DeltaS_A_1st_3D}
\\
\Delta^{2} S_{\rA} {} &= - \tfrac{5}{2} \!\! \int \!\! \rd r \, \frac{[\Delta m' (r)]^{2}}{m' (r)} ,
\label{eq:exp_DeltaS_A_2nd_3D}
\end{align}
\label{eq:exp_DeltaS_A_3D}\end{subequations}
while those of the total energy are given by
\begin{subequations}
\begin{align}
\Delta E_{\rA} {} & = -\tfrac{5}{2}\mG \!\! \int \!\! \rd r \, \frac{m(r)}{r^2}\Delta m(r),
\label{eq:exp_DeltaE_A_1st_3D}
\\
\Delta^{2} E_{\rA} {} & = -\tfrac{5}{2}\mG \!\! \int \!\! \rd r \, \frac{[\Delta m (r)]^{2}}{r^2}.
\end{align}
\label{eq:exp_DeltaE_A_3D}\end{subequations}

As in 1D, we first perform a sanity check of our calculations
by considering the case of a thermodynamical equilibrium,
undergoing adiabatic perturbations.
Imposing the constraints from Eq.~\eqref{eq:def_rT_3D},
we find from Eqs.~\eqref{eq:exp_DeltaS_A_1st_3D} and~\eqref{eq:exp_DeltaE_A_1st_3D}
that, at first order, the variations in the total entropy and total energy
are connected via
\begin{equation}
\Delta S_{\rA} \big|_{\rT} = \frac{3}{2 u_{0}} \, \Delta E_{\rA} \big|_{\rT} ,
\label{eq:link_DSA_DSE_thrm_3D}
\end{equation}
As could have been expected, this relation is identical to the one obtained
in Eq.~\eqref{eq:equality_DS_DE_3D}, where we considered arbitrary perturbations.

We now consider the second-order variation of the entropy.
Using the same approach as in Eq.~\eqref{eq:def_SAtot},
we can compute the total entropy variation at second order,
while imposing the conservation of the total energy at second order.
Using Eqs.~\eqref{eq:exp_DeltaS_A_3D} and~\eqref{eq:exp_DeltaE_A_3D},
we obtain
\begin{equation}
\Delta S_{\rA , \rtot} \big|_{\rT} = - \tfrac{5}{4} \!\! \int \!\! \rd r \, \bigg[ \frac{[\Delta m' (r)]^{2}}{m'(r)} - \tfrac{3}{2} \frac{\mG}{u_{0}} \frac{[\Delta m(r)]^{2}}{r^{2}} \bigg] .
\label{eq:DeltaSAtot_3D_final}
\end{equation}
As expected, Eq.~\eqref{eq:DeltaSAtot_3D_final}
is strikingly similar to Eq.~\eqref{eq:DeltaStot_normal_therm}
obtained when considering arbitrary perturbations.
In particular, up to a global (positive) prefactor,
both expressions match if one assumes ${ \Delta u (r) \!=\! 0 }$
in Eq.~\eqref{eq:DeltaStot_normal_therm}.
Following Eq.~\eqref{eq:link_rho_m},
one has ${ m' \!\propto\! \rho \!>\! 0}$,
so that the first term in Eq.~\eqref{eq:DeltaSAtot_3D_final} is clearly negative.
Yet, as previously, the overall sign of the full expression
remains not obvious.
We do not push this investigation further here
and refer to~\cite{padmanabhan1989}.

We now move to the case of systems submitted to adiabatic
(and spherically symmetric) perturbations.
In that view, let us therefore consider 3D homogeneous self-gravitating
systems in hydrostatic equilibrium.
Phrased differently, we consider systems characterised by some $\bm$,
such that
\begin{equation}
\begin{cases}
\displaystyle \rho = \cst & \text{(Homogeneous density)}
\\
\displaystyle p' = \rho g & \text{(Hydrostatic equilibrium --- Eq.~\ref{eq:hydrostatic_3D})} .
\end{cases}
\label{eq:def_rH_3D}
\end{equation}
For such systems, since ${ \rho' \!=\! 0 }$,
we readily find from Eq.~\eqref{eq:exp_DeltaS_A_1st_3D}
that the first-order variation of the entropy vanishes.
In addition, recalling from Eq.~\eqref{eq:link_rho_m}
that ${ m' \!\propto\! \rho \!>\! 0 }$,
we find that the second-order variation of the entropy
is negative. As a consequence, we therefore have
\begin{equation}
\Delta S_{\rA} \big|_{\rH} = 0 ;
\quad
\Delta^{2} S_{\rA} \big|_{\rH} < 0 .
\label{eq:conclusion_DeltaA_H_3D}
\end{equation}
This is the main result of this section.
It states that 3D homogeneous systems
in hydrostatic equilibrium are entropy extrema
with respect to (spherically symmetric) adiabatic perturbations,
i.e.\ perturbations that maintain the hydrostratic equilibrium.
This is another example of thermodynamic blocking.
All the discussion elements presented
in the 1D case (after Eq.~\ref{eq:conclusion_DeltaA_H})
also apply to the 3D case.\footnote{Except for the derivation based
on barycenter conservation (Appendix~\ref{app:barycenter}),
since, in 3D, spherically symmetric perturbations do not drive any change
in the barycenter.}
As a final remark, we point out that the conclusion from Eq.~\eqref{eq:conclusion_DeltaA_H_3D}
would still hold if one was to replace the hydrostatic condition
from Eq.~\eqref{eq:hydrostatic_3D}
with some arbitrary relation, ${ p' \!=\! p' [\bm] (r) }$.
However, such a constraint would not be physically motivated,
hence of limited relevance.
Indeed, kinetic theory teaches us that it is the hydrostatic equilibrium
that is to be maintained during relaxation
(see the beginning of Section~\ref{sec:tb}).

\section{Discussion and conclusion}
\label{sec:discussion}

In this work we used thermodynamics to investigate self-gravitating systems. We successfully reproduced the well-known result that thermodynamic equilibria are entropy maxima,
thus validating the formalism. The main results of this paper are Eqs.~\eqref{eq:conclusion_DeltaA_H} and \eqref{eq:conclusion_DeltaA_H_3D}, which show that the variation of the thermodynamic entropy of a self-gravitating system is zero if it is homogeneous and submitted to adiabatic perturbations,
i.e.\ perturbations that alway comply with the hydrostatic equilibrium.
These entropy extrema were also shown to be maxima. 

The present derivation relies heavily
on the existence of a self-consistent relation between the profiles
$\bu$ and $\bg$ (resp.\ $\bp$ and $\bm$) in the 1D (resp.\ 3D) case. 
Such a functional relation (be it implicit or explicit) between the fields defining
the total entropy in Eq.~\eqref{eq:def_generic_S} (resp.\ Eq.~\ref{eq:def_S_3D}) is indeed essential.
Mathematically, they define a "line" in the function space of physical states,
the line of hydrostatic equilibria.
We have shown that homogeneous systems
are entropy maxima along this particular line.

Physically, the present result implies that the redistribution of matter and energy cannot increase the (thermodynamic) entropy of a homogeneous self-gravitating system
if hydrostatic equilibrium must be maintained. Phrased differently,
in such homogeneous self-gravitating distributions,
matter and heat do not flow.
We stress however that this does not imply that entropy does not grow
at all in these systems.
Indeed, placing ourselves within a thermodynamical formalism,
we assumed that $S$ was a functional of only two moments
of the phase space distribution.
As such, the thermodynamic entropy 
depends only on the matter and kinetic energy densities,
namely ${ \rho \!=\! \!\int\! \rd v f }$ and ${\rho T \!=\! \!\int\! \rd v f v^2 }$.
Of course, the full dynamical entropy depends on infinitely more higher moments,
e.g.\@, ${ \!\int\! \rd v f v^3 }$, etc.
Such moments do not have a direct macroscopic meaning
and hence are out of the scope of thermodynamics.
Phrased differently, although the present \textit{thermodynamic} entropy ${ S[\bg , \bu] }$
is frozen in homogeneous systems,
the \textit{kinetic} entropy ${ S [f(x,v)] }$ is allowed to grow.
We now conclude by listing a few possible venues for future works.

\textit{Extended entropies}. In the present work, we limited
ourselves to the usual thermodynamic entropy that involves
only two moments of the phase space distribution function,
namely the matter density and the kinetic energy density.
It would be interesting to investigate improved versions
of thermodynamics, in which entropy
is expressed by more, but still finitely many,
moments of the phase space distribution function.
In particular, one should determine whether the present
thermodynamic blocking also holds within such an extended framework.

\textit{Second-order variations}. Throughout the present work,
when considering thermodynamical equilibria,
we repeatedly faced the difficulty of proving
that their second-order entropy variation is negative,
in particular in Eqs.~\eqref{eq:deltaStot}, \eqref{eq:DStot_therm_adiab}, \eqref{eq:DeltaStot_normal_therm} and~\eqref{eq:DeltaSAtot_3D_final}.
Taking inspiration from the method leveraged in~\cite{padmanabhan1989},
it would be rewarding to put these proofs on firmer grounds.
In particular, one should investigate changes
that arise when restricting oneself to adiabatic perturbations.

\textit{Beyond spherically symmetric perturbations}. In Section~\ref{sec:3d}, when considering 3D systems,
we imposed spherical symmetry to the allowed perturbations.
Naturally, it would be enlightening to go beyond this assumption.
In practice, we expect for the thermodynamical blocking
to still hold in that more generic case,
though the involved computations should prove more cumbersome.

\textit{Numerical exploration in 1D}. The dynamics of systems
driven by 1D gravity can be exactly integrated
(up to round-off errors) using an event-driven scheme~\cite{Noullez+2003}.
It would therefore be interesting
to perform (very) long-term simulations of finite-$N$ 1D homogeneous systems,
to quantify how much their self-consistent relaxation
toward the thermodynamical equilibrium is delayed.
Such simulations would be challenging
given that typical (inhomogeneous) quasi-stationary systems
already require extensive integration times
to reach their thermodynamical equilibrium~\citep{Joyce+2010}.
To prevent the accumulation of too large round-off errors,
such a numerical exploration would likely require
the use of high-precision floating-point computations.

\textit{Numerical exploration in 3D}. Similarly,
one should also investigate numerically the practical impact
of thermodynamic blocking in the relaxation
of 3D homogeneous spheres. This was already considered
in~\cite{Sellwood2015}, which emphasised that
(i) homogeneous spheres undergo more coherent
and longer-lasting fluctuations than their inhomogeneous
counterparts; (ii) individual particles in homogeneous spheres
typically diffuse on a timescale of order ${ N^{-1/2} \Tdyn }$,
with $\Tdyn$ the dynamical time,
while in inhomogeneous spheres,
individual particles diffuse on a timescale
of order ${ N^{-1} \Tdyn }$;
but (iii), the overall relaxation of homogeneous spheres,
as a whole, is greatly delayed compared to the one
of inhomogeneous spheres.
Naturally, one ought to revisit,
both numerically and analytically,
these different trends
in the light of the process of thermodynamic blocking.

\textit{2D gravity}. In the present work,
we limited ourselves to presenting the thermodynamic blocking
in 1D and 3D. We expect for this process to also occur
in 2D self-gravitating systems.
Given that 2D gravity shares deep connexions with the dynamics of 2D point vortices~\citep[see, e.g.\@,][]{Chavanis+2007,Bouchet+2012,Chavanis2023},
it would be worthwhile to explore further the impact
of thermodynamic blocking in the hydrodynamic context.
In that case, one may rely on efficient symplectic methods
to perform robust numerical simulations~\cite[see, e.g.\@,][]{Zhang+1993,SanMiguel2006}.

\textit{Deviations away from homogeneity}. To derive
the thermodynamic blocking, we assumed exact homogeneity.
In practice, one may wonder
how much a given system may differ from homogeneity
for the thermodynamic blocking to still hold.
Phrased differently, one should determine
how wide is the domain of near homogeneous systems
whose dynamics remains greatly affected
by the thermodynamic blocking.
This should prove particularly important for finite-$N$ systems,
in which Poisson fluctuations are unavoidable,
i.e.\ in which exact homogeneity is never truly satisfied.

\vspace{6pt}

\authorcontributions{Barnab{\'a}s Deme:
Conceptualization (lead);
Methodology (lead)
Formal analysis (lead);
Investigation (lead);
Methodology (lead); 
Writing – original draft preparation (lead);
Writing – review \& editing (equal).
Jean-Baptiste Fouvry:
Conceptualization (supporting);
Methodology (supporting);
Formal analysis (supporting);
Investigation (supporting);
Methodology (supporting);
Writing – original draft preparation (supporting);
Writing – review \& editing (equal)}

\funding{This work is partially supported by the grant SEGAL
ANR-19-CE31-0017 and BEYOND-BL ANR-25-CE57-2626
of the French Agence Nationale de la
Recherche.}

\dataavailability{No new data were generated.}

\conflictsofinterest{The authors declare no conflict of interest.} 

\appendixtitles{yes}
\appendixstart
\appendix

\section[\appendixname~\thesection]{Energy functional}
\label{app:energy}

\subsection[\appendixname~\thesection]{1D energy}
\label{app:energy_1D}

In this Appendix, we present
a heuristic derivation of the total energy of a 1D self-gravitating system. For the kinetic (or thermal) energy,
following the definition from Eq.~\eqref{eq:uhalfT}
we have
\begin{align}
\Ekin {} & = \!\! \int \!\! \rd x \, \rho(x) \, u (x)
\nonumber
\\
 {} & = - \frac{1}{2 \mG} \!\! \int \!\! \rd x \, g '(x) \, u(x) .
\label{eq:calc_Ekin}
\end{align}

To compute the 1D gravitational energy, one ought to be careful
since the 1D gravitational interaction potential
does not decay to zero for infinite separation,
contrary to the 3D case.
We start from Poisson equation (Eq.~\ref{eq:poisson}).
Integrating both sides over the domain ${ [-x , x] }$, we find
\begin{equation}
m (x) = - g (x) / \mG ,
\label{eq:link_m_1D}
\end{equation}
with ${ m(x) }$ the mass enclosed within that domain.
Now, let us imagine the build-up of a 1D self-gravitating system
by consecutively adding some small mass element, ${ \rd m }$ to it,
at location $x$.
In practice, in order to preserve the symmetry during this process,
we add ${ \rd m / 2 }$ to both the positive and negative edges,
so as to keep the system's centre in ${ x \!=\! 0 }$.
Doing so, we find that the system's total gravitational energy is
\begin{equation}
\Egr = \!\! \int_{0}^{+ \infty} \!\! \frac{\rd m}{2} \, \mG m x - \!\! \int_{0}^{- \infty} \!\! \frac{\rd m}{2} \, \mG m x ,
\label{eq:calc_Egr}
\end{equation}
where the minus sign in the second term accounts for the fact that the gravitational acceleration, $g$,
points in opposite direction on the positive and negative sides.
In addition, for the potential at the edges during this build-up,
we used ${ \phi (x) \!=\! \mG m x }$.
This follows from a further integration of Eq.~\eqref{eq:poisson}.
Using Eq.~\eqref{eq:link_m_1D},
we can now rewrite Eq.~\eqref{eq:calc_Egr} solely as a function of $g$.
We find
\begin{align}
\Egr {} & = \frac{1}{2 \mG} \!\! \int_{-\infty}^{+ \infty} \!\!\!\! \rd x \, g(x) \, g'(x) \, x 
\nonumber
\\
{} & = \frac{1}{2 \mG}\left( \left[ \frac{[g(x)]^{2}}{2}x \right]_{-\infty}^{+ \infty} - \!\! \int_{-\infty}^{+\infty} \!\!\!\!\!\! \rd x \, \frac{[g(x)]^{2}}{2} \right) .
\label{eq:final_Egr}
\end{align}
We note that the boundary term in the last equality is irrelevant when calculating the functional derivatives with respect to $g$.
It is necessary however to lend a finite value to $\Egr$.
Gathering the kinetic energy, $\Ekin$ (Eq.~\ref{eq:calc_Ekin}),
with the gravitational energy, $\Egr$ (Eq.~\ref{eq:final_Egr}),
we recover the expression of the total energy given in Eq.~\eqref{eq:energy}.

\subsection[\appendixname~\thesection]{3D energy}
\label{app:energy_3D}

We now consider the case of a 3D self-gravitating system. For the kinetic energy, we write
\begin{equation}
\Ekin = \!\! \int \!\! \rd \br \, \rho (\br) \, u (\br) .
\label{eq:calc_Ekin_3D}
\end{equation}
Using the law of ideal gas from Eq.~\eqref{eq:ideal_gas_3D},
along with the relation from Eq.~\eqref{eq:link_rho_m},
this becomes
\begin{equation}
\Ekin = \!\! \int \!\! \rd r \, \tfrac{3}{2} \, 4 \pi r^{2} \, p (r) .
\label{eq:final_Ekin_3D}
\end{equation}

To compute the gravitational energy,
we start from
\begin{equation}
\Egr = - \tfrac{1}{2} \!\! \int \!\! \rd \br \rd \btr \, \frac{G \rho (\br) \rho (\btr)}{|\br \!-\! \btr|} ,
\label{eq:calc_Egrav_3D}
\end{equation}
where the factor $\tfrac{1}{2}$ prevents over-counting.
Performing the usual Legendre expansion
of the gravitational potential~\citep{binney2008},
and leveraging spherical symmetry,
this becomes
\begin{align}
\Egr {} & = - \frac{\mG}{2} \!\! \int \!\! \rd r \rd \tr \, 4 \pi r^{2} \, 4 \pi \tr^{2} \, \frac{\rho (r) \rho (\tr)}{\Max[r , \tr]}
\nonumber
\\
{} & = - \frac{\mG}{2} \!\! \int \!\! \rd r \rd \tr \, \frac{m'(r) m'(\tr)}{\Max[r,\tr]} ,
\label{eq:calc_Egrav_3D_next}
\end{align}
where we used the relation from Eq.~\eqref{eq:link_rho_m}.
Since the integrand from Eq.~\eqref{eq:calc_Egrav_3D_next}
is symmetric with respect to ${ r \!\leftrightarrow\! \tr }$,
it suffices to compute the integral from Eq.~\eqref{eq:calc_Egrav_3D_next}
in the domain ${ 0 \!\leq\! \tr \!\leq\! r \!<\! + \infty }$, and double it.
Equation~\eqref{eq:calc_Egrav_3D_next} becomes
\begin{align}
\Egr {} & = - \mG \!\! \int_{0}^{+ \infty} \!\!\!\!\!\! \rd r \, m' (r) \!\! \int_{0}^{r} \!\!\!\! \rd \tr \, \frac{m' (\tr)}{r}
\nonumber
\\
{} & = - \mG \!\! \int_{0}^{+ \infty} \!\!\!\!\!\! \rd r \, \frac{m' (r) m (r)}{r} ,
\label{eq:calc_Egrav_3D_nextnext}
\end{align}
where we used the fact that ${ m (r \!=\! 0) \!=\! 0 }$.
Gathering the kinetic energy, $\Ekin$ (Eq.~\ref{eq:final_Ekin_3D}),
with the gravitational energy, $\Egr$ (Eq.~\ref{eq:calc_Egrav_3D_nextnext}),
we recover the expression of the total energy functional
given in Eq.~\eqref{eq:def_E_3D}.

\section[\appendixname~\thesection]{1D thermodynamic equilibrium}
\label{app:1D_therm_details}

In this Appendix, we provide details on the computations
presented in Section~\ref{sec:Thermo_Eq}.

\subsection[\appendixname~\thesection]{First-order variations}
\label{app:1D_therm_details_1der}

Let us begin with the computation of ${ \delta S / \delta u (x) }$.
We start from Eq.~\eqref{eq:def_generic_S} and write
\begin{align}
\frac{\delta S}{\delta u(x)} {} & = - \frac{1}{2 \mG} \!\! \int \!\! \rd y \, g'(y) \, \frac{\delta s [g'(y) , u(y)]}{\delta u(x)} 
\nonumber
\\
{} & = - \frac{1}{2 \mG} \!\! \int \!\! \rd y \, g'(y) \, \frac{\p s [g'(y) , u(y)]}{\p u (y)} \, \frac{\delta u (y)}{\delta u (x)} .
\label{eq:calc_S}
\end{align}
Using the definition of the specific entropy from Eq.~\eqref{eq:s}
along with the fundamental identity from Eq.~\eqref{eq:fundamental_identity},
we are left with
\begin{equation}
\frac{\delta S}{\delta u (x)} = - \frac{1}{4 \mG} \frac{g'(x)}{u(x)} .
\label{eq:calc_1st_der_S}
\end{equation}
We can proceed similarly for ${ \delta S / \delta g (x) }$.
Manipulations are cumbersome but straightforward.
One finally arrives at
\begin{equation}
\frac{\delta S}{\delta g (x)} = \frac{1}{2 \mG} \bigg( \frac{u'(x)}{2 u(x)} - \frac{g''(x)}{g'(x)} \bigg) ,
\label{eq:calc_2nd_der_S}
\end{equation}
where we used Eq.~\eqref{eq:s}.
Gathering Eqs.~\eqref{eq:calc_1st_der_S} and~\eqref{eq:calc_2nd_der_S}
into Eq.~\eqref{eq:first_DeltaS},
we finally obtain the leading variation of the total entropy
as given in Eq.~\eqref{eq:delta_S_calculated}.

We may perform a similar calculation for ${ \Delta E }$.
One starts from the definition of Eq.~\eqref{eq:energy}
and obtains the expression of the two functional derivatives
\begin{subequations}
\begin{align}
\frac{\delta E}{\delta g(x)} {} & = \frac{1}{2 \mG} \left[ u'(x) - g(x) \right] ,
\label{eq:derE_wrt_g}
\\
\frac{\delta E}{\delta u(x)} {} & = -\frac{1}{2 \mG} \, g'(x) .
\label{eq:derE_wrt_u}
\end{align}
\label{eq:derE_wrt_gu}\end{subequations}
Using these two expressions in the equivalent of Eq.~\eqref{eq:first_DeltaS},
we readily find the leading variation of the total energy
presented in Eq.~\eqref{eq:DeltaE_almost}.

\subsection[\appendixname~\thesection]{Second-order variations}
\label{app:1D_therm_details_2der}

Let us now compute the second-order variations
of the total entropy and energy.
As required by Eq.~\eqref{eq:second_DeltaS},
this asks for the calculation of three functional derivatives.
Differentiating once more Eqs.~\eqref{eq:calc_1st_der_S} and~\eqref{eq:calc_2nd_der_S},
we find
\begin{subequations}
\begin{align}
\frac{\delta^{2} S}{\delta g(y) \delta g(x)} {} & = -\frac{1}{2 \mG} \frac{\p }{\p x} \bigg[ \frac{\deltaD^{\prime} (x \!-\! y)}{g^{\prime} (x)} \bigg] ,
\label{eq:der2S_gg}
\\
\frac{\delta^{2} S}{\delta g(y) \delta u(x)} {} & = -\frac{1}{4 \mG} \frac{1}{u(x)} \, \deltaD^{\prime} (x \!-\! y) ,
\label{eq:der2S_gu}
\\
\frac{\delta^{2} S}{\delta u(y) \delta u(x)} {} & = \frac{1}{4 \mG} \, \frac{g'(x)}{[ u(x) ]^{2}} \, \deltaD (x \!-\! y) .
\label{eq:der2S_uu}
\end{align}
\label{eq:der2S_wrt_gu}\end{subequations}
At this stage, one could be worried that the right-hand side of Eq.~\eqref{eq:der2S_gg}
is not left unchanged by the change ${ x \!\leftrightarrow\! y }$.
This is because a second-order functional derivative like ${ \delta^{2} S / \delta g(x) \delta g(y) }$
is only prescribed up to antisymmetric terms in ${ x \!\leftrightarrow\! y }$.
Such antisymmetric terms will not play any role in practice when evaluating ${ \Delta^{2} S }$,
since in Eq.~\eqref{eq:second_DeltaS},
the second-order derivative is integrated upon the symmetric function ${ \Delta g(x) \Delta g(y) }$.
A similar argument holds for the expression of ${ \delta^{2} S / \delta g(x) \delta u (y) }$.
Injecting Eq.~\eqref{eq:der2S_wrt_gu} into Eq.~\eqref{eq:second_DeltaS},
we find that the second-order variation of the entropy is
\begin{align}
\Delta^{2} S = \frac{1}{2 \mG} \!\! \int \!\! \rd x \, \bigg[ \frac{[\Delta g' (x)]^{2}}{g'(x)} \!-\! \frac{\Delta g'(x) \, \Delta u (x)}{u (x)} \!+\! \frac{1}{2} \frac{g'(x) \, [ \Delta u (x) ]^{2}}{[u(x)]^{2}} \bigg] .
\label{eq:D2_S_continued}
\end{align}
with the notation ${ \Delta g' (x) \!=\! \rd \Delta g (x) / \rd x }$,
and similar for other derivatives.

We may proceed similarly for the second-order variation
of the total energy. Differentiating Eq.~\eqref{eq:derE_wrt_gu} once more,
we find
\begin{subequations}
\begin{align}
\frac{\delta^{2} E}{\delta g (y) \delta g(x)} {} & = - \frac{1}{2 \mG} \, \deltaD (x \!-\! y) ,
\label{eq:der2E_gg}
\\
\frac{\delta^{2} E}{\delta g(y) \delta u(x)} {} & = -\frac{1}{2 \mG} \, \deltaD^{\prime} (x \!-\! y) ,
\label{eq:der2E_gu}
\\
\frac{\delta^{2} E}{\delta u(y) \delta u(x)} {} & = 0 .
\label{eq:der2E_uu}
\end{align}
\label{eq:der2E_wrt_gu}\end{subequations}
Injecting these expressions into the equivalent of Eq.~\eqref{eq:second_DeltaS},
we recover that the second-order variation
of the total energy is
\begin{equation}
\Delta^{2} E = - \frac{1}{2 \mG} \!\! \int \!\! \rd x \, \bigg[ \big[ \Delta g (x) \big]^{2} \!+\! 2 \Delta g' (x) \Delta u(x) \bigg] .
\label{eq:D2_E_continued}
\end{equation}

\section[\appendixname~\thesection]{1D thermodynamic blocking}
\label{app:1D_blck_details}

In this Appendix, we provide details on the computations
presented in Section~\ref{sec:tb}.

The needed functional derivative in Eq.~\eqref{eq:first_DeltaS_A}
follows from Eq.~\eqref{eq:func_S_A}. It reads
\begin{align}
\frac{\delta S_{\rA}}{\delta g (x)} = {} & - \frac{1}{2 \mG} \!\! \int \!\! \rd y \, \bigg[ \frac{\delta s[g(y),g'(y)]}{\delta g (x)} + s [g(y) , g'(y)] \frac{\delta g'(y)}{\delta g(x)} \bigg]
\nonumber
\\
= {} & - \frac{1}{2 \mG} \!\! \int \!\! \rd y \, \bigg[ \frac{\p s[g(y),g'(y)]}{\p g} g'(y) \deltaD (x \!-\! y) - s[g(y),g'(y)] \deltaD^{\prime} (x \!-\! y) 
\nonumber
\\
{} & \quad \quad \quad \quad - g'(y) \frac{\p s[g(y),g'(y)]}{\p g'} \, \deltaD^{\prime} (x \!-\! y) \bigg] .
\label{eq:calc_dSA}
\end{align}
Performing an integration by parts to get rid of the Dirac deltas,
we obtain
\begin{align}
\frac{\delta S_{\rA}}{\delta g(x)} = \frac{1}{2 \mG} \bigg[ {} & g''(x) \, \bigg( 2 \frac{\p s [g(x), g'(x)]}{\p g'} \!+\! g'(x) \frac{\p^{2} s[g(x) , g'(x)]}{\p g'^{2}} \bigg) 
\nonumber
\\
+ \, {} & [g'(x)]^{2} \frac{\p^{2} s[g(x) , g'(x)]}{\p g \p g'} \bigg] .
\label{eq:calc_dSA_next}
\end{align}
We can now use the explicit expression of the specific entropy (Eq.~\ref{eq:s_A})
to further simplify this equation. We obtain
\begin{equation}
\frac{\delta S_{\rA}}{\p g(x)} = - \frac{3}{4 \mG} \frac{g''(x)}{g'(x)} .
\label{eq:calc_grad_SA}
\end{equation}
The second-order functional derivative of Eq.~\eqref{eq:calc_grad_SA}
follows. It reads
\begin{equation}
\frac{\delta^{2} S_{\rA}}{\delta g(x) \delta g(y)} = \frac{3}{4 \mG} \frac{\p }{\p x} \frac{\p }{\p y} \bigg[ \frac{\deltaD (x \!-\! y)}{g'(x)} \bigg] .
\label{eq:calc_grad_SA_2nd}
\end{equation}
Injecting Eqs.~\eqref{eq:calc_grad_SA} and~\eqref{eq:calc_grad_SA_2nd} into Eq.~\eqref{eq:DeltaS_A},
we finally obtain the expression of the first- and second-order variation of the entropy,
as given in Eq.~\eqref{eq:exp_DeltaS_A}.

We proceed similarly to compute the variations
of the "adiabatic" energy. Starting from Eq.~\eqref{eq:def_EA},
we obtain
\begin{subequations}
\begin{align}
\frac{\delta E_{\rA}}{\delta g(x)} {} & = - \frac{3}{4 \mG} \, g(x) ,
\label{eq:calc_grad_EA_1st}
\\
\frac{\delta^{2} E_{\rA}}{\delta g(x) \delta g(y)} {} & = - \frac{3}{4 \mG} \, \deltaD (x \!-\! y) .
\label{eq:calc_grad_EA_2nd}
\end{align}
\label{eq:calc_grad_EA}\end{subequations}
Using the equivalent of Eq.~\eqref{eq:DeltaS_A},
we finally obtain the expression of the first- and second-order
variations of the adiabatic energy,
as given in Eq.~\eqref{eq:exp_DeltaE_A}.

\section[\appendixname~\thesection]{1D thermodynamic blocking from barycenter conservation}
\label{app:barycenter}

In this Appendix, we present an alternative derivation
of the thermodynamic blocking in 1D,
leveraging here the conservation of the system's barycenter.

Let us define the barycenter of a 1D self-gravitating system as
\begin{equation}
X = \!\! \int \!\! \rd x \, \rho(x) \, x
\end{equation}
Leveraging Poisson equation (Eq.~\ref{eq:poisson}),
we can rewrite it as
\begin{equation}
X [\bg] = - \frac{1}{2 \mG} \!\! \int \!\! \rd x \, g'(x) \, x,
\label{eq:def_X}
\end{equation}
We point out that $X$ is independent of the internal energy profile, $\bu$.

Just like for total energy as presented in the main text (see Section~\ref{sec:tb}),
in the absence of external influences,
the system's barycenter must remain constant.
Following the same approach as in Eq.~\eqref{eq:first_DeltaS_A},
the first-order variation of $X$ reads
\begin{align}
\Delta X {} & = -\frac{1}{2 \mG} \!\! \int \!\! \rd x \, \Delta g' (x) x
\nonumber
\\
{} & = \frac{1}{2 \mG} \!\! \int \!\! \rd x \, \Delta g (x) .
\label{eq:barycenter}
\end{align}

Following Section~\ref{sec:tb},
let us now place ourselves in the limit of adiabatic perturbations.
In that case, Eq.~\eqref{eq:exp_DeltaS_A_1st} predicts for the leading-order
variation of the entropy to be
\begin{equation}
\Delta S_{\rA} = - \frac{3}{4 \mG} \!\! \int \!\! \rd x \, \frac{g'' (x)}{g'(x)} \, \Delta g (x) .
\label{eq:rewrite_SA}
\end{equation}
Comparing this expression with Eq.~\eqref{eq:barycenter},
it seems fruitful to consider whether one can devise equilibrium states such that
\begin{equation}
\frac{g'' (x)}{g'(x)} = \alpha ,
\label{eq:alpha_extremum}
\end{equation}
with $\alpha$ a constant.
Indeed, if such systems were to exist,
one would naturally have
\begin{equation}
\Delta S_{\rA} = - \tfrac{3}{2} \, \alpha \, \Delta X = 0 ,
\end{equation}
so that we would have found many more configurations undergoing a thermodynamic blocking.
Indeed, $\alpha$ would be a free parameter,
while only the case ${ \alpha \!=\! 0 }$
corresponds to the case a homogeneous solution,
as uncovered in Section~\ref{sec:tb}.

Upon further scrutiny, at hydrostatic equilibrium,
we expect for the density $\rho$ to be an even function.
Following Poisson equation (Eq.~\ref{eq:poisson}),
$g'$ would then be an even function,
and $g''$ would be an odd one.
Consequently, the function ${ x \!\mapsto\! g''(x) / g'(x)} $
would be odd, so that ${ \alpha \!=\! 0 }$ would be the only viable choice
in Eq.~\eqref{eq:alpha_extremum}.
Imposing ${ \alpha \!=\! 0 }$, we get ${ g'' \!=\! 0 }$ from Eq.~\eqref{eq:alpha_extremum}.
This leads us back to the homogeneous case.
To summarise, leveraging the conservation of the barycenter,
leads to the exact same class of thermodynamic blocking
as obtained in the main text.

To conclude this section, we point out that the present argument,
based on the conservation of the barycentre,
cannot be used in the 3D case. Indeed, when submitted to spherically symmetric perturbations
(as is assumed in Section~\ref{sec:3d}),
the system's barycentre is naturally, and exactly, conserved.

\section[\appendixname~\thesection]{Jeans instability via thermodynamics}
\label{app:Jeans}

In this Appendix, we set out to give a heuristic explanation of the Jeans instability of self-gravitating systems using thermodynamics. To do so, we first assume that perturbations are isothermal, i.e. temperature is kept constant.
In such systems, the most convenient thermodynamic potential to describe equilibrium is the free energy~\cite{atkins2007}. We define it for a 1D self-gravitating system as
\begin{equation}
F [\bg , \bu] = E [\bg , \bu] - \!\! \int \!\! \rd x \, T(x) \, \rho (x) \, s[g'(x) , u(x)] ,
\label{eq:def_F}
\end{equation}
where the energy, ${ E [\bg , \bu] }$, follows from Eq.~\eqref{eq:energy},
the temperature, $T$,
was introduced in Eq.~\eqref{eq:uhalfT},
and the specific entropy, $s$, was given in Eq.~\eqref{eq:s}.
As in usual thermodynamics, the definition from Eq.~\eqref{eq:def_F}
should be interpreted as the Legendre transform
"energy minus temperature times entropy".
We recall that, here, the temperature of the system
is directly given by its internal energy profile, via Eq.~\eqref{eq:uhalfT}.
Expressing $F$ directly as a functional of ${ (\bg , \bu) }$
using Eq.~\eqref{eq:poisson}, we find
\begin{equation}
F [\bg , \bu] = - \frac{1}{2 \mG} \!\! \int \!\! \rd x \, \bigg[ u(x) g'(x) + \tfrac{1}{2} [g(x)]^{2} - 2 u(x) g'(x) \, s [g'(x) , u (x)] \bigg] .
\label{eq:more_explicit_F}
\end{equation}

Let us now perturb this system, through gravity perturbations,
while keeping the temperature fixed.\footnote{Of course, such perturbations are never strictly isothermal. Fortunately, this assumption suffices for a qualitative picture of gravitational instability.}
Phrased differently, we impose some perturbations, ${ \bD \bg }$,
to the system while imposing ${ \bD \bu \!=\! 0 }$.
Following Eq.~\eqref{eq:variations_S_A},
the associated variations in $F$ are then given by
\begin{equation}
F [\bg \!+\! \eps \bD \bg , \bu] \simeq F + \eps \, \Delta F + \tfrac{1}{2} \eps^{2} \Delta^{2} F + ... ,
\label{eq:variations_F}
\end{equation}
with the shortened notations, ${ F \!=\! F [\bg , \bu] }$,
${ \Delta F \!=\! \Delta F [\bg, \bu ; \bD \bg] }$,
and similarly for ${ \Delta^{2} F }$.
Similarly to Eq.~\eqref{eq:DeltaS_A},
under that process, the variations of the free energy
are generically given by
\begin{subequations}
\begin{align}
\Delta F {} & = \!\! \int \!\! \rd x \, \frac{\delta F}{\delta g(x)} \, \Delta g (x)
\label{eq:first_DeltaF}
\\
\Delta^{2} F {} & = \!\! \int \!\! \rd x \, \frac{\delta^{2} F}{\delta g(x) \delta g(y)} \, \Delta g(x) \, \Delta g(y) .
\label{eq:second_DeltaF}
\end{align}
\label{eq:DeltaS_F}\end{subequations}
We may now compute the two functional derivatives in Eq.~\eqref{eq:DeltaS_F}.
Following Eq.~\eqref{eq:more_explicit_F}
and using the results from Eqs.~\eqref{eq:calc_2nd_der_S} and~\eqref{eq:derE_wrt_g},
we find that the first functional derivative is given
\begin{equation}
\frac{\delta F}{\delta g(x)} = -\frac{1}{2 \mG} \bigg( -2u'(x)+g(x)+2u'(x)s-2u(x)\frac{g''(x)}{g'(x)} \bigg) .
\label{eq:func_deltaF_1st}
\end{equation}
Taking one additional functional derivative,
we are left with
\begin{align}
\frac{\delta^{2} F}{\delta g(x) \delta g(y)} = {} & - \frac{1}{2 \mG} \, \deltaD (x \!-\! y) \, + \frac{u'(x)\deltaD'(x-y)}{\mG g'(x)}
\nonumber
\\
{} & + \frac{1}{\mG} \, u (x) \, \bigg[ \frac{\deltaD'' (x \!-\! y)}{g'(x)}-\frac{g''(x)\deltaD'(x-y)}{g'^2(x)} \bigg] .
\label{eq:func_deltaF_2nd}
\end{align}

We may now investigate the stability of infinite homogeneous media
of uniform temperature. As such, we consider states characterized by ${ (\bg , \bu) }$
such that
\begin{equation}
\begin{cases}
\displaystyle \rho = \rho_{0} = \cst & \text{(Constant density)} ,
\\
\displaystyle u = u_{0} = \cst & \text{(Constant temperature --- Eq.~\ref{eq:uhalfT})} .
\end{cases}
\label{eq:def_rC}
\end{equation}
Injecting Eq.~\eqref{eq:func_deltaF_1st} into Eq.~\eqref{eq:first_DeltaF},
we find that the first-order variation of the free energy is given by
\begin{equation}
\Delta F \big|_{\rC} = - \frac{1}{2 \mG} \!\! \int \!\! \rd x \, \bigg[ g(x) - 2 u_{0} \frac{g'' (x)}{g'(x)} \bigg] \, \Delta g (x) .
\label{eq:exp_DeltaF_1st}
\end{equation}
where the subscript "$|_{\rC}$" emphasizes that we are computing
variations around a state with constant density and temperature.
If the system is in hydrostatic equilibrium,
Eq.~\eqref{eq:link_ugp} imposes ${ 2 u_{0} g'' \!=\! g g' }$.
Although, strictly speaking, this cannot be satisfied in a system of infinite extent,
we assume nonetheless that this is correct.
This trick is known as Jeans swindle~\citep[see, e.g.\@, section~{5.2.2} in][]{binney2008}.
Imposing the hydrostatic equilibrium in Eq.~\eqref{eq:exp_DeltaF_1st}
leads to
\begin{equation}
\Delta F \big|_{\rC} = 0 .
\label{eq:vanish_DeltaF_1st}
\end{equation}

The variation in the free energy is therefore of second order.
Injecting Eq.~\eqref{eq:func_deltaF_2nd} into Eq.~\eqref{eq:second_DeltaF},
it reads
\begin{equation}
\Delta^{2} F \big|_{\rC} = -\frac{1}{2 \mG} \!\! \int \!\! \rd x \, \bigg[ [ \Delta g (x) ]^{2} + \frac{2 u_{0}}{g'(x)} \, [ \Delta g' (x) ]^{2} \bigg] .
\label{eq:exp_DeltaF_2nd}
\end{equation}
Using Poisson equation (Eq.~\ref{eq:poisson})
along with our assumption of a constant density profile (Eq.~\ref{eq:def_rC}),
we can rewrite Eq.~\eqref{eq:exp_DeltaF_2nd} as
\begin{equation}
\Delta^{2} F \big|_{\rC} = - \frac{1}{2 \mG} \!\! \int \!\! \rd x \, \bigg[ [\Delta g (x)]^{2} + \frac{u_{0}}{\mG \rho_{0}} \, \Delta g (x) \, \Delta g'' (x) \bigg] ,
\label{eq:exp_DeltaF_2nd_again}
\end{equation}
where we performed an integration by parts of the second term,
neglecting boundary terms.

Finally, we assume that the gravitational perturbation
is of the form
\begin{equation}
\Delta g (x) = A \, \cos (\kp x) ,
\label{eq:shape_Deltag_F}
\end{equation}
with $\kp$ a given (inverse) scale and $A$ a constant.
Injecting this perturbation into Eq.~\eqref{eq:exp_DeltaF_2nd_again},
we find finally
\begin{equation}
\Delta^{2} F \big|_{\rC} = - \frac{A^{2}}{2 \mG} \bigg[ 1 \!-\! \bigg( \frac{\kp}{\kJ} \bigg)^{2} \, \bigg] \! \int \!\! \rd x \, [ \cos (k x) ]^{2} ,
\label{eq:final_DeltaF}
\end{equation}
where we introduced the (inverse) Jeans scale
\begin{equation}
\kJ = \sqrt{ G \rho_{0} / u_{0}} .
\label{eq:def_kJ}
\end{equation}
From Eq.~\eqref{eq:final_DeltaF}, we conclude
that for ${ \kp \!>\! \kJ }$, i.e. for perturbations
on scales small enough,
one has ${ \Delta^{2} F |_{\rC} \!>\! 0 }$,
i.e.\ the system is stable to perturbations.
On the contrary, for ${ \kp \!<\! \kJ }$,
i.e.\ for perturbations on scales larger than the Jeans scale,
one has ${ \Delta^{2} F |_{\rC} \!<\! 0 }$,
i.e.\ the system is unstable with respect to such perturbations.

We note that this line of arguments lacks rigor for several reasons. First, the variation in \eqref{eq:shape_Deltag_F} does not vanish at the boundaries.
This is a key assumption in thermodynamics.
Second, the integral in Eq.~\eqref{eq:final_DeltaF},
although clearly positive, is not convergent
in a system of infinite extent.
All these limitations are directly connected to the Jeans swindle.
Despite these limitations,
Eq.~\eqref{eq:final_DeltaF} can serve as a useful guide
to understand the role played by thermodynamics
in self-gravitating systems.

\section[\appendixname~\thesection]{3D thermodynamic equilibrium}
\label{app:3D_thrm_details}

In this Appendix, we provide details on the computations
presented in Section~\ref{sec:3d:TE}.

\subsection[\appendixname~\thesection]{First-order variations}
\label{app:3D_thrm_details_1der}

We begin with the computation of ${ \delta S / \delta u (r) }$.
We start with Eq.~\eqref{eq:def_S_3D} and write
\begin{align}
\frac{\delta S}{\delta u(r)} {} & = \!\! \int \!\! \rd \tr \, \tfrac{3}{2} \, m' (\tr) \frac{\deltaD (r \!-\! \tr)}{u (\tr)}
\nonumber
\\
{} & = \tfrac{3}{2} \frac{m'(r)}{u(r)} .
\label{eq:calc_dSdp_3D}
\end{align}
Similarly, for ${ \delta S / \delta m(r) }$, we get
\begin{align}
\frac{\delta S}{\delta m(r)} {} & =  \!\! \int \!\! \rd \tr \, \deltaD' (r \!-\! \tr) \, \bigg[ \tfrac{3}{2} \ln [u(\tr)] -  \ln [\rho (\tr)] \bigg] - \!\! \int \!\! \rd \tr \, m ' (\tr) \,  \, \frac{\deltaD' (r \!-\! \tr) }{4 \pi \tr^{2} \rho (\tr)}
\nonumber
\\
{} & = \frac{\rho' (r)}{\rho(r)} - \tfrac{3}{2} \frac{u' (r)}{u (r)} ,
\label{eq:calc_dSdm_3D}
\end{align}
where we used the relation from Eq.~\eqref{eq:link_rho_m}.
Gathering Eqs.~\eqref{eq:calc_dSdp_3D} and~\eqref{eq:calc_dSdm_3D}
into Eq.~\eqref{eq:DeltaS_3D}, we obtain the leading-order variation
of the entropy, as given in Eq.~\eqref{eq:DeltaS_3D_generic}.

We can perform similar calculations for ${ \Delta E }$.
Starting from the definition of Eq.~\eqref{eq:def_E_3D},
we readily obtain the first functional derivative
\begin{equation}
\frac{\delta E}{\delta u (r)} = \!\! \int \!\! \rd \tr \, \deltaD(r \!-\! \tr) \, m'(\tr) = m'(r) .
\label{eq:calc_dEdp_3D}
\end{equation}
For the other functional derivative, we write
\begin{align}
\frac{\delta E}{\delta m(r)} {} & =   \!\! \int \!\! \rd \tr \, \bigg\{  \deltaD'(r \!-\! \tr) u(\tr)- \mG \bigg[ \deltaD (r \!-\! \tr) \, \frac{m' (\tr)}{\tr} - \deltaD' (r \!-\! \tr) \, \frac{m(\tr)}{\tr} \bigg] \bigg\}
\nonumber
\\
{} & = - \!\! \int \!\! \rd \tr \, \deltaD (r \!-\! \tr)\left(u'(\tilde{r})+ \frac{\mG m(\tr)}{\tr^{2}}\right)
\nonumber
\\
{} & = -u'(r)- \mG \, \frac{m (r)}{r^{2}} .
\label{eq:calc_dEdm_3D}
\end{align}
Gathering Eqs.~\eqref{eq:calc_dEdp_3D} and~\eqref{eq:calc_dEdm_3D}
in the equivalent of Eq.~\eqref{eq:DeltaS_3D},
we finally obtain the leading order variation of the total energy,
as given in Eq.~\eqref{eq:DeltaE_3D_generic}.

\subsection[\appendixname~\thesection]{Second-order variations}
\label{app:3D_thrm_details_2der}

We may now compute the second-order variations
of the total entropy and total energy in the 3D case.
Following Eq.~\eqref{eq:second_DeltaS_3D}, this requires the computation
of three functional derivatives.
Differentiating once more Eqs.~\eqref{eq:calc_dSdp_3D} and~\eqref{eq:calc_dSdm_3D}, we find
\begin{subequations}
\begin{align}
\frac{\delta^{2} S}{\delta u (\tr) \delta u (r)} {} & = - \tfrac{3}{2} \frac{m'(r)}{[u(r)]^{2}} \, \deltaD (r \!-\! \tr) ,
\label{eq:calc_d2Sdp2_3D}
\\
\frac{\delta^{2} S}{\delta m (\tr) \delta u (r)} {} & = \tfrac{3}{2} \frac{\deltaD' (r \!-\! \tr)}{u (\tr)} ,
\label{eq:calc_d2S_dpdm_3D}
\\
\frac{\delta^{2} S}{\delta m (\tr) \delta m (r)} {} & = \frac{\deltaD''(r \!-\! \tr)}{m'(r)}-\frac{m''(r)\deltaD'(r \!-\! \tr)}{[m'(r)]^2} .
\label{eq:calc_d2S_dm2_3D}
\end{align}
\label{eq:calc_d2S_3D}\end{subequations}
Injecting Eq.~\eqref{eq:calc_d2S_3D} into Eq.~\eqref{eq:second_DeltaS_3D},
we find that the second-order variation of the entropy reads
\begin{align}
\Delta^{2} S = \!\! \int \!\! \rd r \rd \tr \, \bigg\{ {} & - \tfrac{3}{2}\frac{m'(r)}{[u(r)]^2} \deltaD(r \!-\! \tr) \Delta u(r)\Delta u(\tr) + 3 \frac{\deltaD'(r \!-\! \tr)}{u(r)}\Delta u(r)\Delta m(\tr)
\nonumber
\\
{} & + \bigg[ \frac{\deltaD''(r \!-\! \tr)}{m'(r)} \!-\! \frac{m''(r)\deltaD'(r \!-\! \tr)}{[m'(r)]^2} \bigg] \Delta m(r)\Delta m(\tr) \bigg\} .
\label{eq:wrap_D2S_3D}
\end{align}

We can proceed similarly for the second-order variation
of the total energy. Differentiating Eqs.~\eqref{eq:calc_dEdp_3D} and~\eqref{eq:calc_dEdm_3D}
once more, we find
\begin{subequations}
\begin{align}
\frac{\delta^{2} E}{\delta u(\tr) \delta u(r)} {} & = 0 ,
\label{eq:calc_d2Edp2_3D}
\\
\frac{\delta^{2} E}{\delta m(\tr) \delta u(r)} {} & = \deltaD'(r \!-\! \tr) ,
\label{eq:calc_d2E_dpdm_3D}
\\
\frac{\delta^{2} E}{\delta m(\tr) \delta m(r)} {} & = - \frac{\mG}{r^{2}} \, \deltaD (r \!-\! \tr) .
\label{eq:calc_d2E_dm2}
\end{align}
\label{eq:calc_d2E_3D}\end{subequations}
Injecting these expressions into the equivalent of Eq.~\eqref{eq:second_DeltaS_3D},
we find that the second-order variation of the energy reads
\begin{equation}
\Delta^{2} E = - \!\! \int \!\! \rd r \rd \tr \, \bigg\{ -\frac{\mG \deltaD(r \!-\! \tr) }{r^2} \Delta m(r) \Delta m(\tr) + 2 \deltaD'(r \!-\! \tr)\Delta u(r) \Delta m(\tr) \bigg\} .
\label{eq:wrap_D2E_3D}
\end{equation}

\section[\appendixname~\thesection]{3D thermodynamic blocking}
\label{app:3D_blck_details}

In this Appendix, we provide details on the computations
presented in Section~\ref{sec:thrm_blck_3D}.
The needed functional derivative in Eq.~\eqref{eq:first_DeltaS_A_3D}
follows from Eq.~\eqref{eq:def_SA_3D}.
To compute it, we need to recall that
we have ${ p(r) \!=\! p [\bm] (r) }$ (via Eq.~\ref{eq:hydro_3D_expl})
and ${ \rho (r) \!=\! \rho [\bm] (r) }$ (via Eq.~\ref{eq:link_rho_m}).
We write
\begin{align}
\frac{\delta S_{\rA}}{\delta m(r)} = \!\! \int \!\! \rd \tr \, \bigg\{ {} & - \deltaD'(r \!-\! \tr)\left[ \tfrac{3}{2} \ln [p(\tr) - \tfrac{5}{2} \ln [\rho(\tr) ] \right]
\nonumber
\\
{} & + m' (\tr) \left[\tfrac{3}{2}\frac{1}{p(\tr)} \frac{\delta p (\tr)}{\delta m(r)} - \tfrac{5}{2} \frac{1}{\rho (\tr)} \frac{\delta \rho (\tr)}{\delta m(r)} \right] \bigg\} .
\label{eq:calc_dS_3D_details}
\end{align}
In order to perform this integration, we note that both ${ r \!\mapsto\! m(r) }$ and ${ r \!\mapsto\! p(r) }$
are bijective functions.
As such, we can formally redefine the pressure as
${ p (m) \!=\! p (r[m]) }$,
where ${ r[m] }$ follows from the inversion of the function
${ r \!\mapsto\! m(r) }$.
The functional relation between $p$ and $m$ is set by
the hydrostatic equilibrium from Eq.~\eqref{eq:hydro_3D_expl}.
It becomes
\begin{equation}
   \frac{\rd p (m)}{\rd m}=-\frac{\mG m}{4\pi [r(m)]^4}.
\end{equation}
Using this expression, we get  
\begin{equation}
    \frac{\delta p(\tr)}{\delta m(r)}=\frac{\rd p (m)}{\rd m}\frac{\delta m(\tr)}{\delta m(r)}=-\frac{\mG m}{4\pi [r(m)]^4}\deltaD(r-\tr).
\end{equation}
Similar arguments hold for the density, where the functional relation between $\rho$ and $m$ is given by Eq.~\eqref{eq:link_rho_m}. The associated functional derivative reads
\begin{equation}
    \frac{\delta \rho(\tr)}{\delta m(r)}=\frac{\rd \rho (m')}{\rd m'}\frac{\delta m'(\tr)}{\delta m(r)}=\frac{1}{4\pi \tr^2}\deltaD'(r-\tr).
\end{equation}
Equipped with these expressions, the functional derivative of the entropy can now be directly computed.
From Eq.~\eqref{eq:calc_dS_3D_details}, one obtains
\begin{equation}
\frac{\delta S}{\delta m(r)} = \tfrac{5}{2}\frac{\rho'(r)}{\rho(r)}.
\label{eq:dSdm_bck_3D_final}
\end{equation}
The second-order functional derivative of Eq.~\eqref{eq:dSdm_bck_3D_final} follows.
It reads
\begin{equation}
\frac{\delta^2 S_{\rA}}{\delta m(r) \delta m(\tr)} = \tfrac{5}{2} \bigg( \frac{\deltaD''(r \!-\! \tr)}{m'(r)} - \frac{m''(r)\deltaD'(r \!-\! \tr)}{m'(r)^2} \bigg),
\end{equation}
Following Eq.~\eqref{eq:second_DeltaS_A_3D},
we finally obtain the second-order variation of the entropy, namely
\begin{align}
\Delta^2 S_{\rA} {} & = \!\! \int \!\! \rd r \rd \tr \, \tfrac{5}{2}\left( \frac{\deltaD''(r \!-\! \tr)}{m'(r)}-\frac{m''(r)\deltaD'(r \!-\! \tr)}{m'(r)^2} \right)\Delta m(r)\Delta m (\tr)
\nonumber
\\
{} & = - \tfrac{5}{2} \!\! \int \!\! \rd r \, \frac{[ \Delta m' (r) ]^2}{m'(r)} ,
\end{align}
as given in Eq.~\eqref{eq:exp_DeltaS_A_2nd_3D}.

We proceed similarly to compute the variations
of the "adiabatic" energy from Eq.~\eqref{eq:def_EA_3D}.
We get
\begin{align}
\frac{\delta E_{\rA}}{\delta m(r)} {} & = \!\! \int \!\! \rd \tr \, \bigg\{ \tfrac{3}{2} 4\pi \tr^2 \frac{\delta p (\tr)}{\delta m (r)}-\frac{\mG\deltaD (r \!-\! \tr)m'(\tr)}{\tr} - \frac{\mG m(\tr) \deltaD (r \!-\! \tr)}{\tr} \bigg\}.
\nonumber
\\
{} & = - \tfrac{5}{2}\frac{\mG m(r)}{r^2}.
\label{eq:calc_dEdm_blck_3D}
\end{align}
Differentiating once more, we finally obtain
\begin{equation}
\frac{\delta^{2} E_{\rA}}{\delta m(r) \delta m(\tr)} = -\tfrac{5}{2}\frac{\mG \deltaD (r \!-\! \tr)}{r^2} .
\label{eq:calc_d2Edm2_blck_3D}
\end{equation}

\reftitle{References}

\end{document}